\documentclass[aps,pra,onecolumn,groupedaddress,longbibliography,showkeys,showpacs]{revtex4-2}

\usepackage{latexsym}
\usepackage{amsmath}
\usepackage{amsfonts}
\usepackage{graphicx}
\usepackage{mathptmx}      
\usepackage{bm}
\usepackage{enumerate}
\usepackage{color}
\definecolor{DarkGreen}{RGB}{0,100,0}

\usepackage{bm}

\newcommand\bb{\mathbf{b}}
\newcommand\be{\mathbf{e}}
\newcommand\bx{\mathbf{x}}

\newcommand\bp{\mathbf{p}}

\newcommand\bB{\mathbf{B}}
\newcommand\bE{\mathbf{E}}

\newcommand\bF{\mathbf{F}}
\newcommand\bJ{\mathbf{J}}
\newcommand\bv{\mathbf{v}}
\newcommand\bs{\mathbf{s}}
\newcommand\bS{\mathbf{S}}
\newcommand\bL{\mathbf{L}}

\newcommand\bM{\mathbf{M}}
\newcommand\bR{\mathbf{R}}

\newcommand\uu{u}
\newcommand\vv{v}
\newcommand\ww{w}

\begin{document}


\title{Classical, quantum and event-by-event simulation of a Stern-Gerlach experiment with neutrons}

\author{H. De Raedt}
\affiliation{Institute for Advanced Simulation, J\"ulich Supercomputing Centre,
Forschungszentrum J\"ulich, D-52425 J\"ulich, Germany\\}
\affiliation{Zernike Institute for Advanced Materials,\\
University of Groningen, Nijenborgh 4, NL-9747 AG Groningen, The Netherlands}
\author{F. Jin}
\affiliation{Institute for Advanced Simulation, J\"ulich Supercomputing Centre,\\
Forschungszentrum J\"ulich, D-52425 J\"ulich, Germany}
\author{K. Michielsen}
\affiliation{Institute for Advanced Simulation, J\"ulich Supercomputing Centre,\\
Forschungszentrum J\"ulich, D-52425 J\"ulich, Germany\\}
\affiliation{RWTH Aachen University, 52056 Aachen, Germany}
\date{\today}

\begin{abstract}
We present a comprehensive simulation study of the Newtonian and quantum model of a
Stern-Gerlach experiment with cold neutrons.
By solving Newton's equation of motion and the time-dependent Pauli equation ,
for a wide range of uniform magnetic field strengths,
we scrutinize the role of the latter for drawing the conclusion that
the magnetic moment of the neutron is quantized.
We then demonstrate that a marginal modification of the Newtonian
model suffices to construct, without invoking any concept of quantum theory,
an event-based subquantum model that eliminates the shortcomings of the classical model and yields
results that are in qualitative agreement with experiment and quantum theory.
In this event-by-event model, the intrinsic angular momentum can take any value on the sphere,
yet, for a sufficiently strong uniform magnetic field, the particle beam splits in two,
exactly as in experiment and in concert with quantum theory.
\end{abstract}


\maketitle

\section{Introduction}\label{INTRO}

In 1922, O. Stern and W. Gerlach demonstrated experimentally that silver atoms
passing through an inhomogeneous magnetic field experience deflections in
spatially different, distinguishable directions. This observation was very important
for the early development of quantum theory for it provided direct experimental
evidence that not only the spectra of atoms but also the magnetic moment of the
particles might be quantized~\mbox{\cite{STER22,Frisch1933,Gerlach1924}}. The Stern--Gerlach (SG)
experiment is often used in textbooks~\cite{BOHM51,FEYN65,BAYM74,BALL03} to
introduce the concepts of spin and quantization of angular momentum and plays a
prominent role in discussions on determining properties of atomic size objects
by means of macroscopic measuring devices~\cite{Hannout1998,
SchmidtBoecking2019}. The SG experiment, and its conceptually equivalent
experiment with single photons passing through a birefringent crystal, are also
used in textbooks to illustrate postulates of quantum
theory~\cite{BOHM51,FEYN65,BAYM74,BALL03}.

\begin{figure}[!htp]

\includegraphics[width=0.9\hsize]{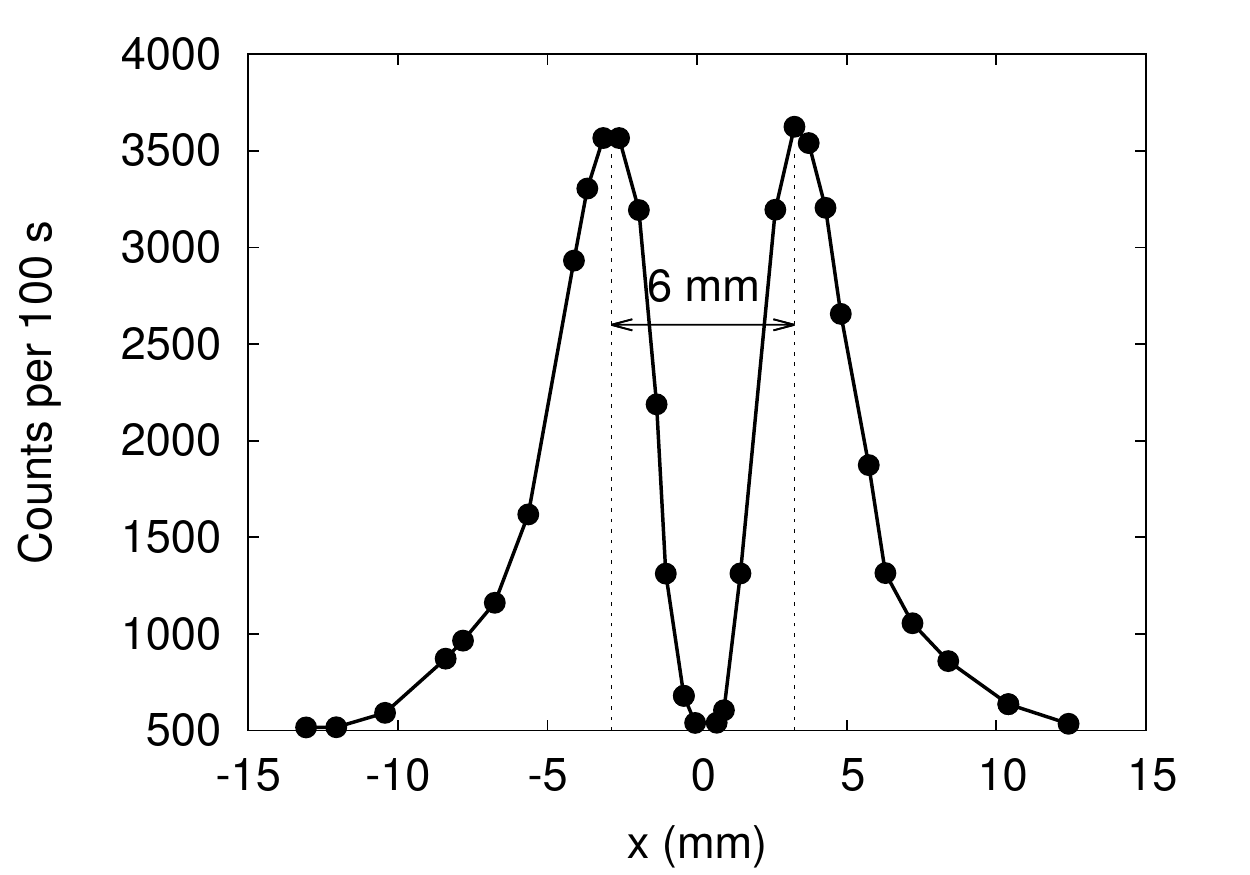}
\caption{
Diagram (not to scale) of a Stern--Gerlach experiment with cold neutrons,
performed by Hamelin et al.~\cite{HAME75}.
After passing through the collimators, most neutrons travel along the $y$-direction.
The cone indicates the directions in which the neutron may, but not necessarily have to,
leave the magnetic field region.
\label{EXPI0}
}
\end{figure}

In short, an SG experiment involves a source of electrically neutral, magnetic particles,
collimators, a magnet generating an inhomogeneous field, and a particle detector;
see Figure~\ref{EXPI0} for a sketch of an SG with cold neutrons. Due to the
interaction between the magnetic moment of the particle and the inhomogeneous
magnetic field, a particle passing through the latter experiences a force that
changes the trajectory of the particle. Note that this reasoning is entirely
Newtonian, no concept of quantum theory is entering yet.

Assuming (i) uniformly random orientations of the magnetic moments leaving the source and (ii) a
sufficiently large uniform magnetic field, the standard classical picture of the
magnetic moment as a spinning top leads to the conclusion that there should be
no splitting of the beam~\cite{FEYN65}; however, under certain
conditions~\cite{Platt1992}, to be scrutinized in the present paper, the SG
magnet splits the particle beam in two, spatially well-separated directions, in
agreement with the outcome of the SG experiment. As the amount of deflection is
proportional to the magnetic moment, an SG-like apparatus can be
used to measure the magnetic moment of nano-size particles~\cite{Heer1990,BACH18}.

As originally conceived, the SG experiment employs electrically neutral
particles. Obviously, this begs the question if it would be feasible
to perform a similar experiment to observe the spin of say,
electrons~\cite{BATE97,RUTH98,BATE98a,Garraway1999} or ions~\cite{Henkel2019}.
Addressing this interesting question is beyond the scope of
the present paper, which focuses on the case of electrically neutral particles only.

The deflection in spatially well-separated directions along the direction of the
uniform magnetic field is commonly regarded as an experimental proof that the
magnetic moment of the particles is
quantized~\cite{STER22,Frisch1933,BOHM51,FEYN65}. Labeling the distinct beams by
a two-valued variable $s=\pm1/2$ and representing the beams by the corresponding
state vectors forms the basis for the well-known quantum-theoretical description
of the idealized SG experiment~\cite{BOHM51,FEYN65,BAYM74,BALL03,RAED18a,RAED19b}.

The first aim of the present paper is primarily pedagogical
in that we present, to the best of our knowledge, the first comprehensive treatment
of both the Newtonian and quantum model of a real SG experiment. In order to
touch base with a real SG experiment, we have taken model parameters from an SG
experiment performed with cold neutrons~\cite{HAME75}. In this respect, there is
little overlap with earlier numerical studies of the quantum model of an SG
experiment~\cite{POTE05,Hsu2011}.

The second aim is to demonstrate that a minor modification to the classical,
Newtonian equations of motion in the spirit of the event-by-event simulation
approach yields results that (i) can be very different from those of the classical
and (ii) are in full qualitative agreement with SG experiments and
with the quantum-theoretical description thereof.
The idea behind this modification is the following.
As long as the particle does not experience a magnetic field,
the internal frame of reference used to define the direction of the magnetic moment
is detached from the laboratory frame of reference.
This hold true in quantum theory as well: in the absence of an electromagnetic field
there is no relation between the $xyz$-coordinates of the particle and $xyz$-components
of the spin operator~\cite{BALL03}.
In the event-based approach, a particle
moving from a field-free region into a region where the electromagnetic field is present
is viewed as an event which establishes the relation between the
$xyz$-coordinates of the particle and $xyz$-components of the magnetic moment.
This event-triggered process of alignment may be thought of as a highly simplified model
for the classical electrodynamic transient processes that occur when
a magnetic moment moves through a region in which the magnetic field changes~\cite{JACK62}.

The paper is structured as follows.
Section~\ref{EXPI} describes the SG experiment with neutrons~\cite{HAME75} that
we take as reference for our simulation work.
In Sections~\ref{NEWT} and~\ref{QTM}, we present and discuss the results obtained by
solving Newton's equation of motion and the time-dependent Pauli equation (TDPE), respectively.
Adopting the parameters for the cold neutrons SG experiment
in combination with the macroscopic size of the experimental
setup requires the use of high-precision solvers and substantial computer resources.
In Section~\ref{QTM}, we also discuss the transition from
a description in terms of position and spin to a model that involves spin-1/2 operators only.
Section~\ref{EVENT} introduces the modification to Newton's equation of motion that
turns the classical model into a event-by-event, subquantum model for the SG experiment,
meaning that data generated by the latter exhibit the same features
as the data obtained by SG experiments and their quantum-theoretical description.
Section~\ref{CONC} summarizes our findings.

\section{Neutron Experiment}\label{EXPI}

Figure~\ref{EXPI0} shows a schematic of the SG experiment
with neutrons, as performed by Hamelin et al.~\cite{HAME75}.
Cold neutrons leaving the neutron guide
impinge on a collimator positioned $0.2\;\mathrm{m}$
from the exit plane of the neutron guide.
The selected neutrons impinge on a second collimator, placed
$1\;\mathrm{m}$ from the first one.
The strongly collimated beam of neutrons then passed
through the SG magnet which is $0.8\;\mathrm{m}$ long.
The distance between the second collimator and the exit
plane of the SG magnet is $0.9\;\mathrm{m}$.
The direction of the neutrons leaving the SG magnet
is selected by means of a meaning window.
The distance between the exit plane of the SG magnet
and the $^3\mathrm{He}$ detector is $2\;\mathrm{m}$.

\begin{figure}[!htp]

\includegraphics[width=0.9\hsize]{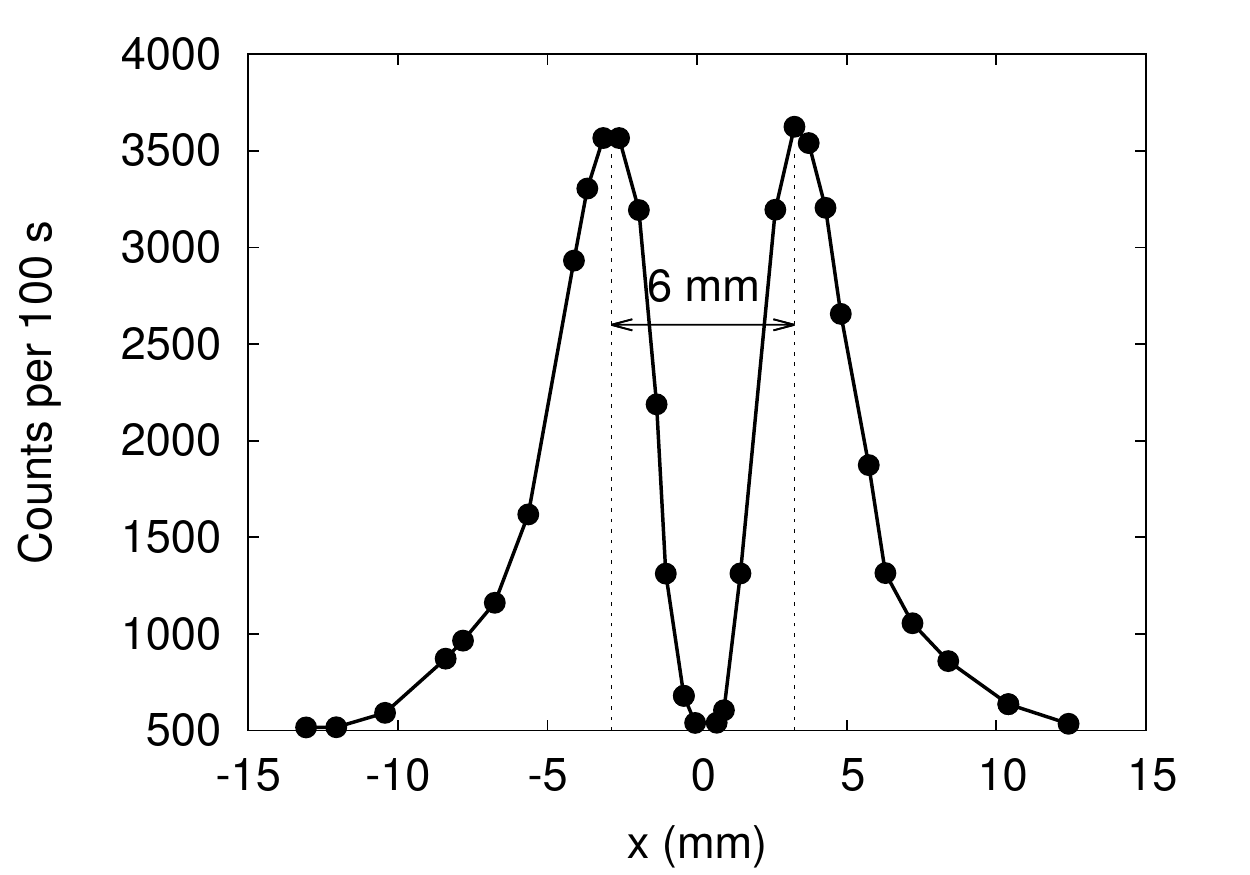}
\caption{{}
Neutron counts per 100 s as recorded in the SG
experiment by \mbox{Hamelin et al.~\cite{HAME75}.}
The data were extracted from Figure~6 of Ref.~\onlinecite{HAME75} by hand.
\label{EXPI1}
}
\end{figure}

In Figure~\ref{EXPI1} we present some of the results reported in Ref.~\onlinecite{HAME75}.
Clearly, the SG magnet causes the neutron beam to split in two well-defined
beams, with their maxima of intensities separated by about $6\;\mathrm{mm}$.
Note that the window (see Figure~\ref{EXPI0}) in front of the detector moves in the
$x$-direction only.

Looking at the experimental data presented in Figure~\ref{EXPI1},
it is obvious that in order to represent the spin state of a neutron
by a two-valued variable, it is necessary to classify the
data points as belonging to one of two groups.
As the two maxima of the counts are well separated, simply drawing
a vertical line at $x=0$ suffices to classify the data points.
Once this classification is made, we can dispose of the spatial degree of freedom
and describe the process of spin-based filtering in terms of spin-1/2 matrices,
a model that is often used in textbooks~\cite{BOHM51,FEYN65,BAYM74, BALL03}.

\section{Newtonian Mechanics}\label{NEWT}
The Hamiltonian describing the dynamics of a neutral, particle of mass $m$ and
magnetic moment $\bM$ subject to a time-independent magnetic field $\bB=\bB(\bx)$ reads
\begin{eqnarray}
H &=& 
\frac{m}{2}{\bv}^2 - \bM\cdot\bB(\bx)\;
=\frac{m}{2}{\bv}^2 - \gamma\,\bL\cdot\bB(\bx)\;,
\label{NEWT0}
\end{eqnarray}
where $\bL$ is the angular momentum relative to the center of mass $\bx$ of the particle,
and $\gamma$ is the gyromagnetic ratio.
Starting from Equation~(\ref{NEWT0}), the standard procedure to derive the equations of motion yields,
%
\begin{eqnarray}
m\frac{d\bv}{d t}& =&\gamma\; \bm\nabla\,(\bB(\bx)\cdot\bL) \;,
\label{NEWT1}
\\
\frac{d\bL}{d t}& =&\gamma\; \bL\times\bB(\bx)\;.
\label{NEWT2}
\end{eqnarray}
The angular momentum $\bL$ has the same dimension as $\hbar$, namely $\left(\mathrm{kg}\; \mathrm{m}^2\; \mathrm{s}^{-1} \right)$.
In order to facilitate the comparison with the quantum-theoretical description, it is expedient to define
$\bL=\hbar \bS$ where
$\bS=S(\cos\phi\sin\theta,\sin\phi\sin\theta,\cos\theta)^{\mathrm{T}}$
is a dimensionless vector.
In terms of this vector, the classical equations of motion read
%
\label{NEWT3}
\begin{eqnarray}
\frac{m}{\hbar}\frac{d\bv}{d t}& =&\gamma\; \bm\nabla\,(\bB(\bx)\cdot\bS) \;,
\label{NEWT3a}
\\
\frac{d\bS}{d t}& =&\gamma\; \bS\times\bB(\bx)\;.
\label{NEWT3b}
\end{eqnarray}
Note that the presence of $\hbar$ is the result of
rewriting the classical equations of motion in
terms of a dimensionless angular momentum $\bS$
and does not, in any way, imply that Equations~(\ref{NEWT3a})--(\ref{NEWT3b}) describes quantum phenomena.
The length $S$ of the vector $\bS$ does not affect the solution of {Equation~(\ref{NEWT3b})}
and needs to be fixed by comparison with the results of the quantum-theoretical description; this is described in a later section of the paper. 

\subsection{Model for the Magnetic Field}\label{MAGF}

Essential to an SG experiment is that the magnetic particles
interact with an {\sl inhomogeneous} magnetic field.
Maxwell's equation requires that $\bm\nabla\cdot\bB(\bx)=0$.
From the Maxwell equation
{\color{black}
\begin{eqnarray}
\frac{\partial \bE(\bx,t)}{\partial t}=\frac{1}{\epsilon\mu}\bm\nabla\times\bB(\bx)-\frac{1}{\epsilon}\bJ(\bx,t)
\;,
\label{QTM1a}
\end{eqnarray}
where $\bJ(\bx,t)$, $\epsilon$ and $\mu$ represent the external current, the electrical permittivity,
and magnetic permeability, respectively.}
It follows that if $\bm\nabla\times\bB(\bx)\not=0$, the magnetic field would induce a nonzero,
time-dependent electric field $\bE(\bx,t)$.
The strength of this electric field would increase linearly with time.
Although this electric field would not affect the motion of the electrically neutral particles,
in our study, we only consider the case $\bm\nabla\times\bB(\bx)=0$.

A simple choice, complying with the  conditions
$\bm\nabla\cdot\bB(\bx)=0$ and $\bm\nabla\times\bB(\bx)=0$
just mentioned,
is~\cite{Majorana1932,Alstrom1982,Scully1987,Platt1992,POTE05,Hsu2011}
\begin{eqnarray}
\bB(\bx)&=&
\left\{\begin{array}{ll}(B_0+zB_1)\be_z-xB_1\be_x&\;,\;y\in[y_0,y_1]\\0&\;,
\;y\not\in[y_0,y_1]\\
\end{array}\right.
\;,
\label{QTM1}
\end{eqnarray}
that is, $\bB(\bx)=0$ except when $y_0\le y \le y_1$ where the strength of the field gradient
in both the $x$ and $z$ direction is $B_1>0$ (we adopt the convention that $B_0,B_1\ge0$).
The term in Equation~(\ref{QTM1}) proportional to $B_0$, the uniform magnetic field in the $z$-direction,
describes the contribution of the dipole field.
The two terms in Equation~(\ref{QTM1}) proportional to $B_1$ are characteristic
for the quadrupole contribution to the magnetic field.
The values of $B_0$ and $B_1$ depend on the design of the magnet.
In this paper, we regard $B_0$ and $B_1$ as model~parameters.

From Equation~(\ref{QTM1}) it follows that for
$y\in[y_0,y_1]$, the force $\bF(\bx)$ on the particle is given~by
\begin{eqnarray}
\bF(\bx)&=&\gamma\;\bm\nabla\,(\bB(\bx)\cdot\bS)=\gamma\;B_1 (S^z\be_z-S^x\be_x)
\;,
\label{NEWT4}
\end{eqnarray}
independent of $x$ or $z$.
For $y\not\in[y_0,y_1]$, the force $\bF(\bx)$ on the particle is zero.
{\color{black}
As a function of $y$, the simple model Equation~(\ref{QTM1}) shows discontinuities at $y=y_0,y_1$.
Instead of smoothing out these discontinuities, we integrate
the equations of motion in the interval $y_0\le y\le y_1$
and assume that the velocity distribution at $y = y_1$
is representative (up to trivial, free-particle scale factors)
for the velocity and position distributions at $y\gg y_1$.
}

From Equation~(\ref{NEWT4}), it follows immediately that the velocity in the $y$-direction is conserved.
In this paper, we assume that all particles move
with velocity $v_y$ along the $y$-direction.
The time it takes for the particles to traverse
the magnetic field region is given by $t^\ast=(y_1-y_0)/v_y$.

Once a particle's $y$-coordinate exceeds $y_1$, its velocity $\bv=(v_x,v_y,v_z)$ is used
to increment the histogram count at the transverse velocity coordinate $(v_x,v_z)$ and the simulation
of that particle is terminated.
The distribution of transverse velocities $(v_x,v_z)$ does not
change if the particles leave the region where the magnetic field is present
and is therefore well-suited to analyze the data.
The distribution of transverse positions $(x,z)$ at any plane located to the right of
the SG (see Figure~\ref{EXPI0}) is straightforwardly obtained
from the distribution of transverse velocities by using the fact
that in the field-free region, the particles propagate freely.

In this paper, we mainly present results for the distribution of the transverse velocities $(v_x,v_z)$,
obtained by classical, quantum-theoretical, and event-by-event simulation.
This distribution contains all information about the outcome of the simulated SG experiment
and facilitates the presentation of the simulation data in a compact, unified, and convenient~manner.

\subsection{Analytically Solvable Cases}\label{ANALY}
It is of interest to consider a special case that is easy to solve analytically.
We take as initial positions and velocities of the $N$ particles
$\bx=(0,y_0,0)$ and $\bv=(0,v_y,0)$, respectively,
and we only consider the case in which all particles have their initial magnetic moment
along the $z$-axis, i.e.,  $\bS=S(0,0,\pm1)$.
Note that $\bS\times\bB(\bx=(0,y,z))=0$ for any $(y,z)$, see Equation~(\ref{QTM1}),
implying that for $\bx=(0,y,z)$, the torque on the spin is zero; therefore, the direction of the spin does not change and
the particles only feel a constant force in the $z$-direction, see Equation~(\ref{NEWT4}).
The trajectory is that of a particle in a constant force field,
that is $v_z(t)=\pm \hbar\gamma B_1 S t/m$ and
$z(t)=\pm \hbar\gamma B_1 S t^2/2m$ for $0\le t \le t^\ast$.

Looking ahead, this simple scenario mimics the quantum-theoretical textbook case (see Appendix~\ref{TEX}) and
allows us to fix the magnitude of the classical magnetic moment $\bS$.
Indeed, the classical and quantum-theoretical expressions
for the change in the velocity due to the magnetic field gradients match if $S=1/2$.

From the analysis of the analytically solvable, classical mechanical case, it follows that
the time of flight, the changes of transverse velocity and displacement are given by
\begin{equation}
t^\ast=\frac{y_1-y_0}{v_y}\;,\;v^\ast= \left|\frac{\hbar\gamma B_1}{2m}\right|t^\ast\;,\;
z^\ast=\frac{v^\ast t^\ast}{2}
\;,
\label{NEWT5}
\end{equation}
respectively.
The three parameters Equation~(\ref{NEWT5}) characterize the state of the particles at
the point $y=y_1$, that is when they leave the region where the magnetic field is present.
Again, looking ahead, the quantum-theoretical textbook case also yields Equation~(\ref{NEWT5}).
We use Equation~(\ref{NEWT5}) to set the scale of time, velocity, and position
for both the classical and quantum-theoretical model.

The second solvable case is the one that is often referred to when
comparing the classical and quantum-theoretical picture of the magnetic moment.

If the uniform magnetic field is present ($B_0>0$), a
transformation to a frame rotating with angular frequency $\gamma B_0$ removes
the static field term $-\gamma B_0 S^z$ from
{\color{black}the transformed Hamiltonian
at the cost of introducing time-dependent, sinusoidal terms in
the equations of motion.}
Then, the argument goes, if these sinusoidal terms oscillate sufficiently
rapidly, their effect on the motion averages out~\cite{BAYM74,Platt1992}.
Although this argument holds for $B_0\rightarrow\infty$,
for realistic values of $B_0$ and $B_1$, see Section~\ref{ESTI}, it does not.
Only if the particle trajectories are close to the region
where the field gradient is small, the argument applies, see Appendix~\ref{NMRa}.
When applied to the SG experiment with realistic values of $B_0$ and $B_1$,
the above argument is circular but self-consistent.
The justification that the argument is valid comes from
the numerical solution presented in Section~\ref{NSIM}.

If we simply omit the $x$-component in Equations~(\ref{QTM1}) and~(\ref{NEWT4}) (and thereby
violate one of Maxwell's equations),
we are left with the classical problem in which $S^z$ does not change
with time and the particle is subject to a force
$\bF(\bx)=\gamma\;B_1 S^z\be_z$ (recall that $B_0$ has disappeared because
of the transformation to the rotating frame).
For the initial conditions $\bx=(0,y_0,0)$ and $\bv=(0,v_y,0)$ we have
$v_x(t^\ast)=0$, $v_z(t^\ast)=\pm \hbar\gamma B_1 S^z t^\ast/m$,
$x(t^\ast)=0$, and $z(t^\ast)=\pm \hbar\gamma B_1 S^z (t^\ast)^2/2m$.
{\color{black}
The expressions for final velocity $v_z(t^\ast)$ and position $z(t^\ast)$
are the same as those obtained in the first analytically solvable case.}
For each random choice of $\bS$, $S^z$ is a random number in the range $[-1/2,1/2]$
and the distribution of velocities is a line at $v_x=0$, stretching from $v_z=-v^\ast$ to  $v_z=v^\ast$.
This is the expected outcome of the Newtonian description of the SG experiment
that is often referred to when comparing with the quantum-theoretical prediction.

\subsection{Model Parameters}\label{ESTI}

We adopt the geometry of the experiment with neutrons, reported in Ref.~\onlinecite{HAME75}.
The region in which there is a nonzero gradient
in the $x$-$z$ directions is $0.8\,\mathrm{m}$ long~\cite{HAME75},
that is $y_1-y_0=0.8\,\mathrm{m}${\color{black}. In} the neutron experiment, the maximum gradient of the $B$-field is
estimated to be {\color{black}$B_1= 300\; \mathrm{T}/\mathrm{m}$}~\cite{HAME75}.
In the case of the SG experiment with silver atoms, estimates range from
$B_1=1\; \mathrm{T}/\mathrm{cm}=100\; \mathrm{T}/\mathrm{m}$ to
$B_1=20\; \mathrm{T}/\mathrm{cm}=2000\; \mathrm{T}/\mathrm{m}$~\cite{SchmidtBoecking2019,Vigue2019}.
In view of the uncertainties about the strength and precise form of the $B$-field
gradients in these experiments and taking into consideration that
the simple form of the $B$-field gradients used for
our theoretical/simulation study is unlikely to hold to any of these experiments,
we will use $B_1= 300\; \mathrm{T}/\mathrm{m}$ in all our simulation work.

In the case of the experiment with neutrons we have~\cite{RAUC15,HAME75}
{\color{black}
\begin{eqnarray}
\begin{array}{lclclclc}
m&=&1.67\times10^{-27}\,\mathrm{kg}
&,&
\gamma&=&-1.83 \times10^{8}\,\mathrm{T}^{-1}\mathrm{s}^{-1}
&,
\\
|\gamma B_0|&=&1.83\times10^{8}\,\mathrm{s}^{-1}
&,&
|\gamma B_1|&=&5.50\times10^{10}\,\mathrm{m}^{-1}\,\mathrm{s}^{-1}
&,
\\
{\hbar B_1}/{m B_0}&=&1.89\times10^{-5}\,{\mathrm{m}}\,{\mathrm{s}}^{-1}
&,&
{\hbar|\gamma| B_1}/{m}&=&{3.46\times10^{3}}\,{\mathrm{m}}\,{\mathrm{s}^{-2}}
&,
\\
v_y&=&395.6\,{\mathrm{m}}\,{\mathrm{s}}^{-1}
&,&
t^\ast&=&2.02\times10^{-3}\,\mathrm{s}
&,
\\
v^\ast&=&3.50\,{\mathrm{m}}\,{\mathrm{s}}^{-1}
&,&
z^\ast&=&3.53\times10^{-3}\,\mathrm{m}
&,
\end{array}
\label{VALU0}
\end{eqnarray}
}
where $\hbar=1.05\times10^{-34}\,\mathrm{kg}\,\mathrm{m}^{2}\mathrm{s}^{-1}$ and
we have taken as an example $B_0=1\,\mathrm{T}$.

Assume, as we did in Section~\ref{ANALY}, that all particles have
their initial magnetic moment along the $z$-axis, i.e.,  $\bS=S(0,0,\pm1)$.
According to Equation~(\ref{NEWT5}), the particles cross the plane at $y=y_1$
at $\bx=(0,y_1,\pm v^\ast t^\ast/2)=(0,y_1,\pm 3.53\times10^{-3}\;\mathrm{m})$
with a velocity $\bv=(0,v_y,\pm3.50\,\mathrm{m}\,\mathrm{s}^{-1})$.
During the remaining free-particle flight to the detector screen,
the $z$-coordinate changes by $\Delta z_{\mathrm{screen}} =
\pm 3.50 \times2 \,\mathrm{m}/395.6=\pm 17.7\,\mathrm{mm}$.
Thus, in traveling from the source to the detector,
the $z$-coordinate changes by $\Delta z_{\mathrm{source-screen}}\approx \pm 21.2\,\mathrm{mm}$.
This is about a factor of 7 larger than the splitting observed in the neutron
experiment~\cite{HAME75}; see Figure~\ref{EXPI0}.
In view of the fact that the magnetic field Equation~(\ref{QTM1}) is unlikely to result
from the real magnet used in the experiment~\cite{HAME75}, this order-of-magnitude
agreement between the beam-splittings at the screen is quite satisfactory.

{\color{black}
As explained in Appendix~\ref{appQTM}, solving the time-dependent Pauli equation for the quantum-theoretical
model with the set of parameters given by Equation~(\ref{VALU0}) is computationally very expensive.
In order to speed up the development of the simulation software
and to generate simulation data for a case that is substantially different than that of neutrons,
we have chosen to perform simulations with
parameters taken from the original SG experiment~\cite{SchmidtBoecking2019,Vigue2019}
except that instead of the value of magnetic moment of the silver atom,
we have taken the value of the magnetic moment of the Ag$^{107}$ nucleus~\cite{Brun1954}.
In the following, we refer to this case as simulations with imaginary silver particles.
The parameters are
\begin{eqnarray}
\begin{array}{lclclclc}
m&=&1.79\times10^{-25}\,\mathrm{kg}
&,&
\gamma&=&-1.09 \times10^{7}\,\mathrm{T}^{-1}\mathrm{s}^{-1}
&,
\\
|\gamma B_0|&=&1.09\times10^{7}\,\mathrm{s}^{-1}
&,&
|\gamma B_1|&=&3.26\times10^{9}\,\mathrm{m}^{-1}\,\mathrm{s}^{-1}
&,
\\
{\hbar B_1}/{m B_0}&=&1.76\times10^{-7}\,{\mathrm{m}}\,{\mathrm{s}}^{-1}
&,&
{\hbar|\gamma| B_1}/{m}&=&1.91\,{\mathrm{m}}\,{\mathrm{s}^{-2}}
&,
\\
v_y&=&540\,{\mathrm{m}}\,{\mathrm{s}}^{-1}
&,&
t^\ast&=&1.48\times10^{-3}\,\mathrm{s}
&,
\\
v^\ast&=&1.42\times10^{-3}\,{\mathrm{m}}\,{\mathrm{s}}^{-1}
&,&
z^\ast&=&1.05\times10^{-6}\,\mathrm{m}
&,
\end{array}
\label{VALU1}
\end{eqnarray}
where again, we have taken as an example $B_0=1\,\mathrm{T}$.
}

\subsection{Numerical Solution of Equation~(\ref{NEWT3})}\label{XXXX}

In practice, we solve the system Equation~(\ref{NEWT3}) by a combination of the exact integration
of the torque equation Equation~(\ref{NEWT3b}) and the velocity-Verlet method as used in molecular dynamics~\cite{RAPA04}.
Appendix~\ref{VECTOR} gives the details of the algorithm that we use.

Unless mentioned explicitly, the model parameters for all our classical simulations are
$B_1=300\, \mathrm{T}/\mathrm{m}$.
Numerical experiments show that the simulation results show insignificant quantitative
changes if we decrease the time step from $\tau=10^{-8} ~\mathrm{s}$ to
$\tau=10^{-9} ~\mathrm{s}$. We use the latter to compute the data that we present in this paper.

Solving Equation~(\ref{NEWT3}) for $N=$ 1,000,000 particles
with a time step of $\tau=10^{-9}\,\mathrm{s}$ takes of the order of hundred minutes
on a compute node with two 24-cores Intel Xeon Platinum 8168 CPUs running at 2.7 GHz.
We only present data that are essential for the comparison
of the classical and quantum description of an SG experiment.

\subsection{Newtonian Dynamics: Simulation Results for Neutrons}\label{NSIM}

\begin{figure}[!htp]

\includegraphics[width=0.47\hsize]{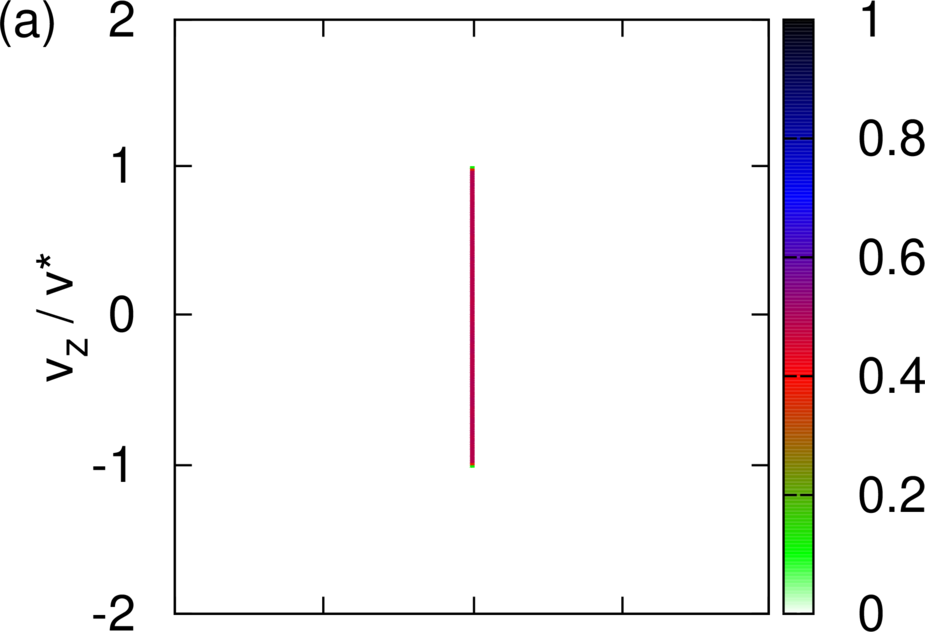}
\hspace{0.022\hsize}%
\includegraphics[width=0.47\hsize]{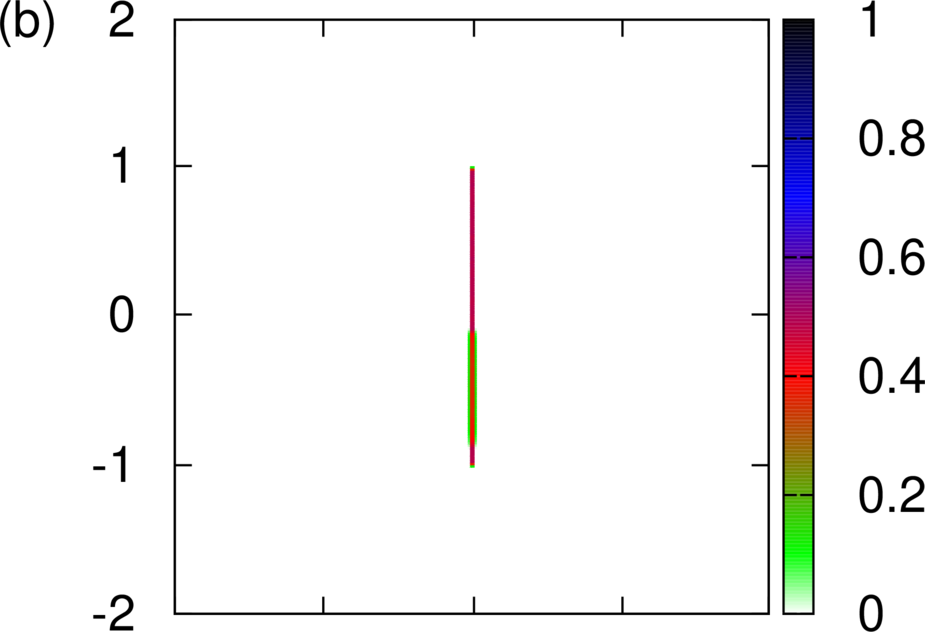}\\
\vspace{0.022\hsize}%
\includegraphics[width=0.47\hsize]{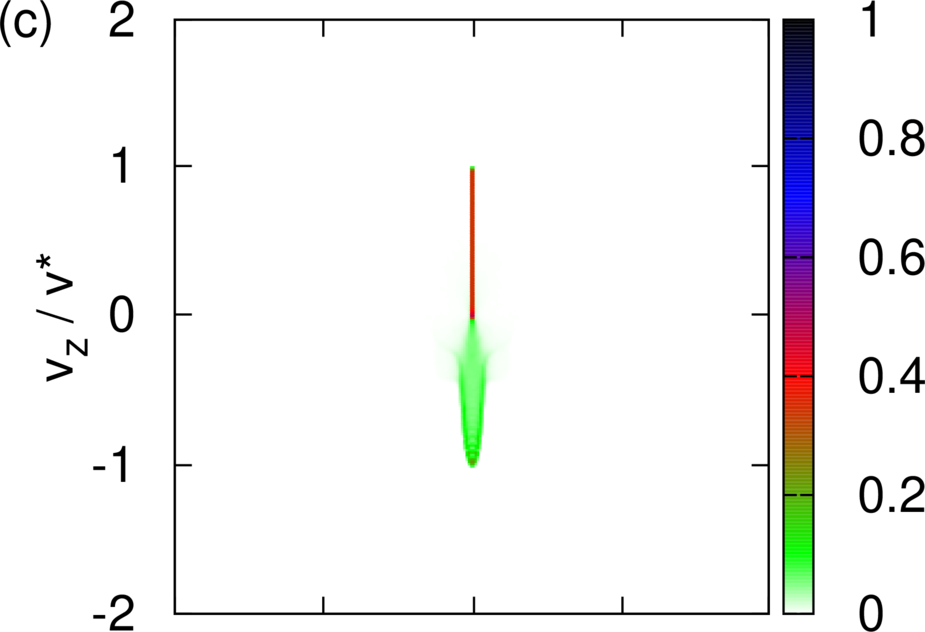}
\hspace{0.022\hsize}%
\includegraphics[width=0.47\hsize]{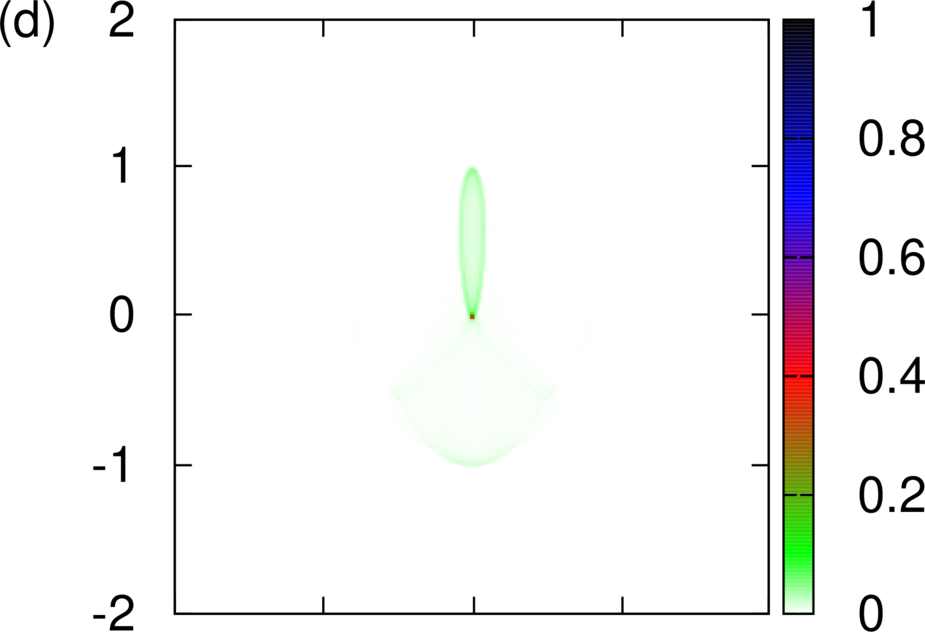}\\
\vspace{0.022\hsize}%
\includegraphics[width=0.47\hsize]{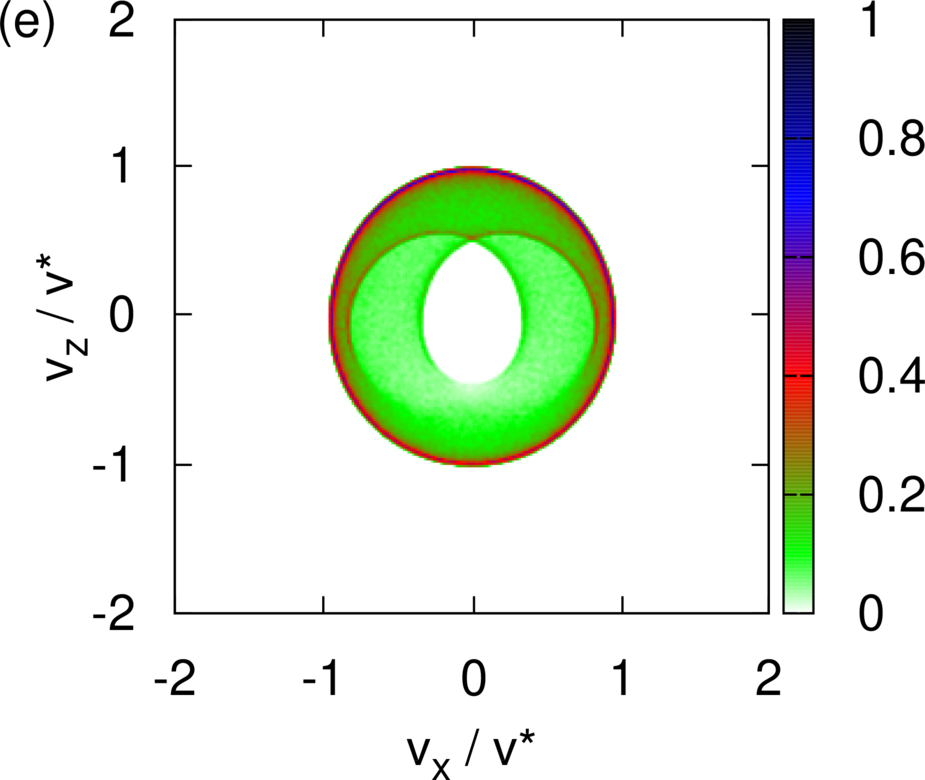}
\hspace{0.022\hsize}%
\includegraphics[width=0.47\hsize]{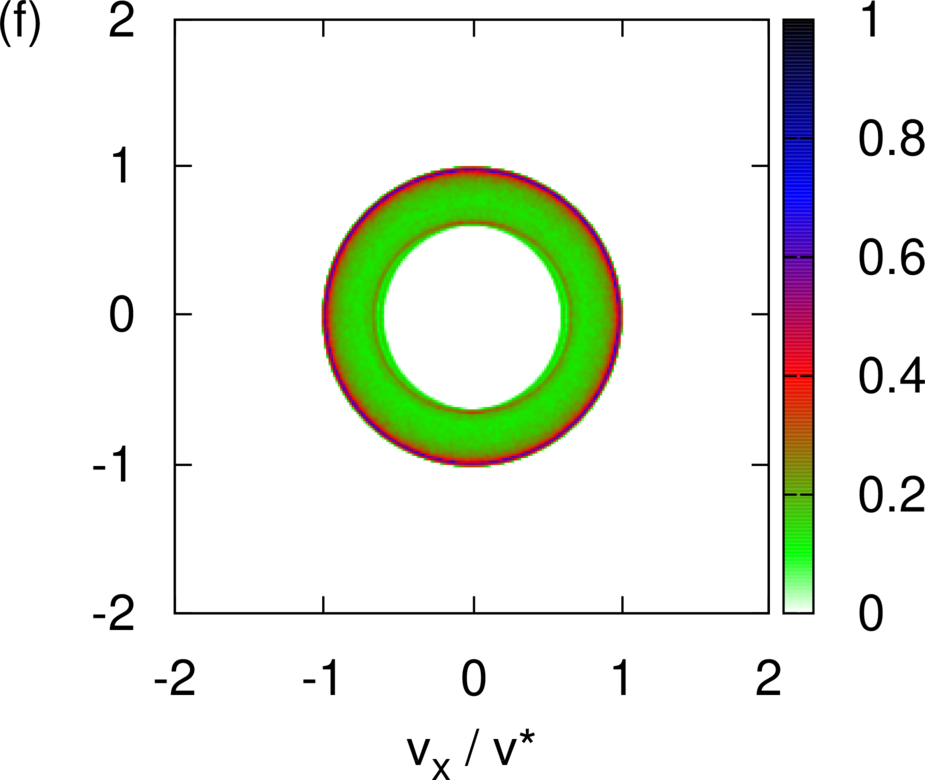}
\caption{{}
Histograms of the transverse velocity distribution obtained by
the solving the classical equations of motion Equation~(\ref{NEWT3})
with the initial magnetic moments distributed randomly (see text)
and for different values of the uniform magnetic field $B_0$.
(\textbf{a}) $B_0=1\; \mathrm{T}$;
(\textbf{b}) $B_0=0.1\; \mathrm{T}$;
(\textbf{c})~$B_0=0.01\; \mathrm{T}$;
(\textbf{d}) $B_0=0.001\; \mathrm{T}$, hard to see but looks similar to a projection of an elongated pacifier;
(\textbf{e})~$B_0=0.0001\; \mathrm{T}$;
(\textbf{f}) $B_0=0.00001\; \mathrm{T}$.
\label{NSIM0}
}
\end{figure}

{In this section, we focus on the SG with neutrons~\cite{HAME75}.
Repeating the simulations with the particle parameters of {\color{black}imaginary silver particles} (see Equation~(\ref{VALU1})
yields data, some of which are presented in {{Figure~\ref{NSIM0}} and} Appendix~\ref{NSIMAG}, that leads to the
same general~conclusions.}

To allow for a spreading of the particle beam entering the magnet,
the distance from the source to the magnet $y_0=1\; \mathrm{m}$,
similar to the distance between the rightmost collimator and the magnet in the neutron experiment~\cite{HAME75}.
The length of the magnet in the $y$-direction $y_1-y_0=0.8\; \mathrm{m}$~\cite{HAME75}.
The distance from the magnet to the detection screen is taken to be zero because the motion
of the particle is that of a free particle with velocity $\bv=(\Delta v_x,v_y, \Delta v_z)$
where $\Delta v_x$ and $\Delta v_z$ are the changes of the transverse velocities
due to the magnetic field~gradients.

As the particles leave the source, the positions and velocities are
normally distributed, centered  around $\bx=(0,0,0)$ and $\bv=(0,v_y,0)$
and with variances $\sigma_x$ and $\sigma_v$, respectively.
Unless mentioned explicitly, $\sigma_x=\sigma_v=0$.

The simulation reproduces the analytically obtained results if the initial positions and velocities are
$\bx=(0,y_0,0)$ and $\bv=(0,v_y,0)$, respectively,
and the initial magnetic moments are aligned along the $z$-axis, i.e.,  $\bS=(0,0,\pm1)/2$.
The results are in excellent agreement with those obtained by solving the problem analytically
and are, therefore, not shown.

Next, we assume that the direction of the magnetic moment,
represented by the three-dimensional spin vector $\bS$,
is uniformly distributed over the sphere.
In different words, there is maximum uncertainty about the
directions of magnetic moments of the neutrons emerging from the neutron guide (see Figure~\ref{EXPI0}).
With this initial condition of $\bS$, the transverse velocity distribution
changes drastically as the strength of the uniform magnetic field decreases
from rather strong ($B_0=1\,\mathrm{T}$) to very weak ($B_0\approx0\,\mathrm{T}$),
as illustrated in Figure~\ref{NSIM0}a--f.

From Figure~\ref{NSIM0}a, it follows that a uniform magnetic field of $B_0=1\; \mathrm{T}$
is sufficiently strong to suppress the effect of the $x$-component of the magnetic field.
Because the spins $\bS$ of different particles are distributed uniformly over the sphere of radius $S=1/2$,
the final distribution of velocities is a strip at $v_x\approx0$, stretching from $v_z=-v^\ast$ to  $v_z=v^\ast$.
This is exactly as expected~\cite{STER22,GERL24,FEYN65,BOHM51,BAYM74,BALL03} on the basis of the arguments
discussed in Section~\ref{ANALY}.

Figures~\ref{NSIM0}b--e demonstrate that the transverse velocity distribution changes
drastically each time we reduce $B_0$ by an order of magnitude.
Analytically predicting any of particular shapes shown in Figure~\ref{NSIM0}c--e
seems to be a daunting task.

For $B_0=0$, any rotation of the $(x,z)$ coordinates
about the $y$-axis together with the corresponding inverse rotation of the spin leaves the Hamiltonian invariant.
As the initial values of $(S^x,S^z)$ are distributed uniformly over a circle
it follows that the maxima of the transverse velocity distribution
are expected to trace out a circle in the $v_x$-$v_z$ plane, in agreement
with Figure~\ref{NSIM0}f.

\begin{figure}[!htp]

\includegraphics[width=0.40\hsize]{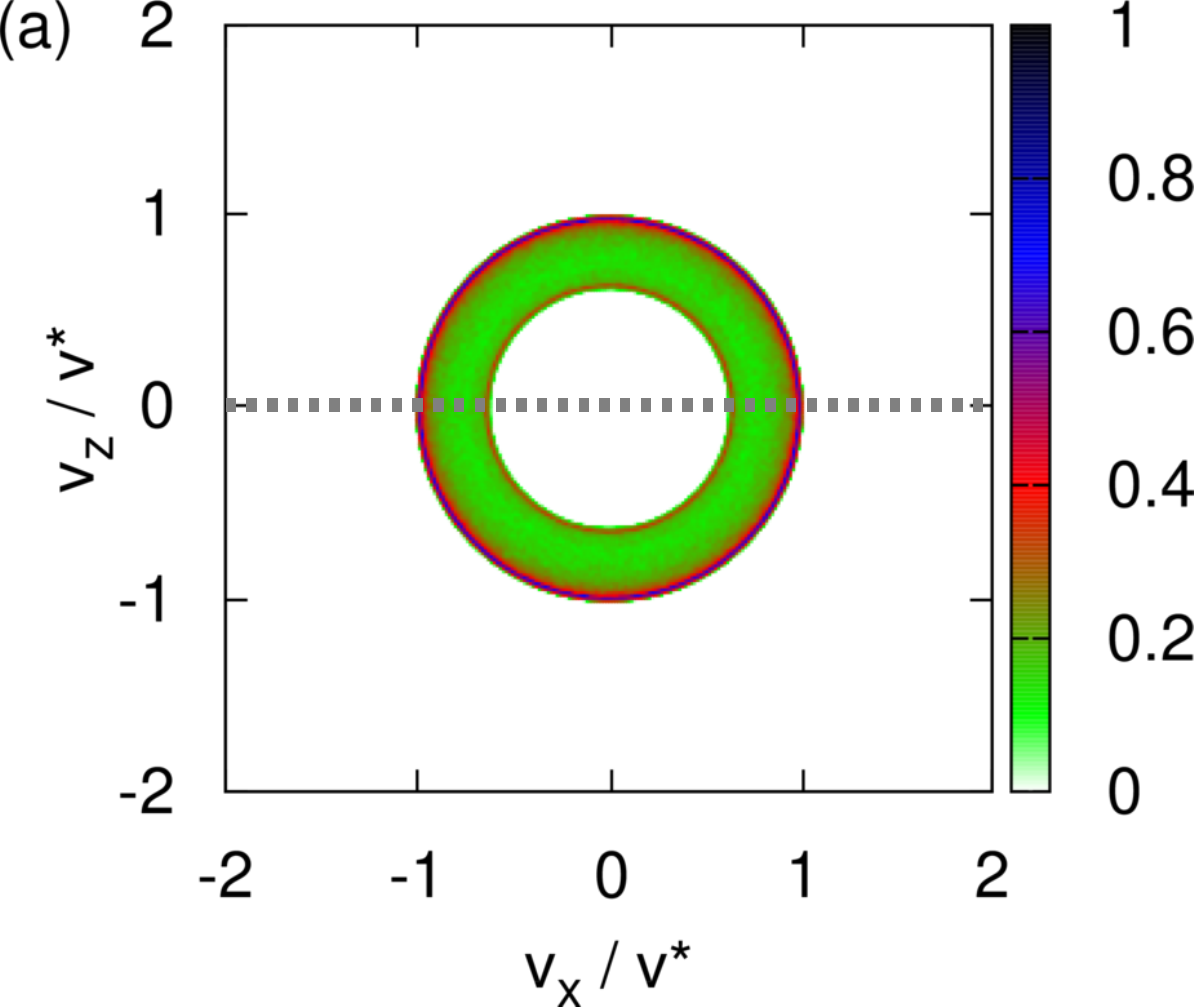}
\hspace{0.022\hsize}%
\includegraphics[width=0.55\hsize]{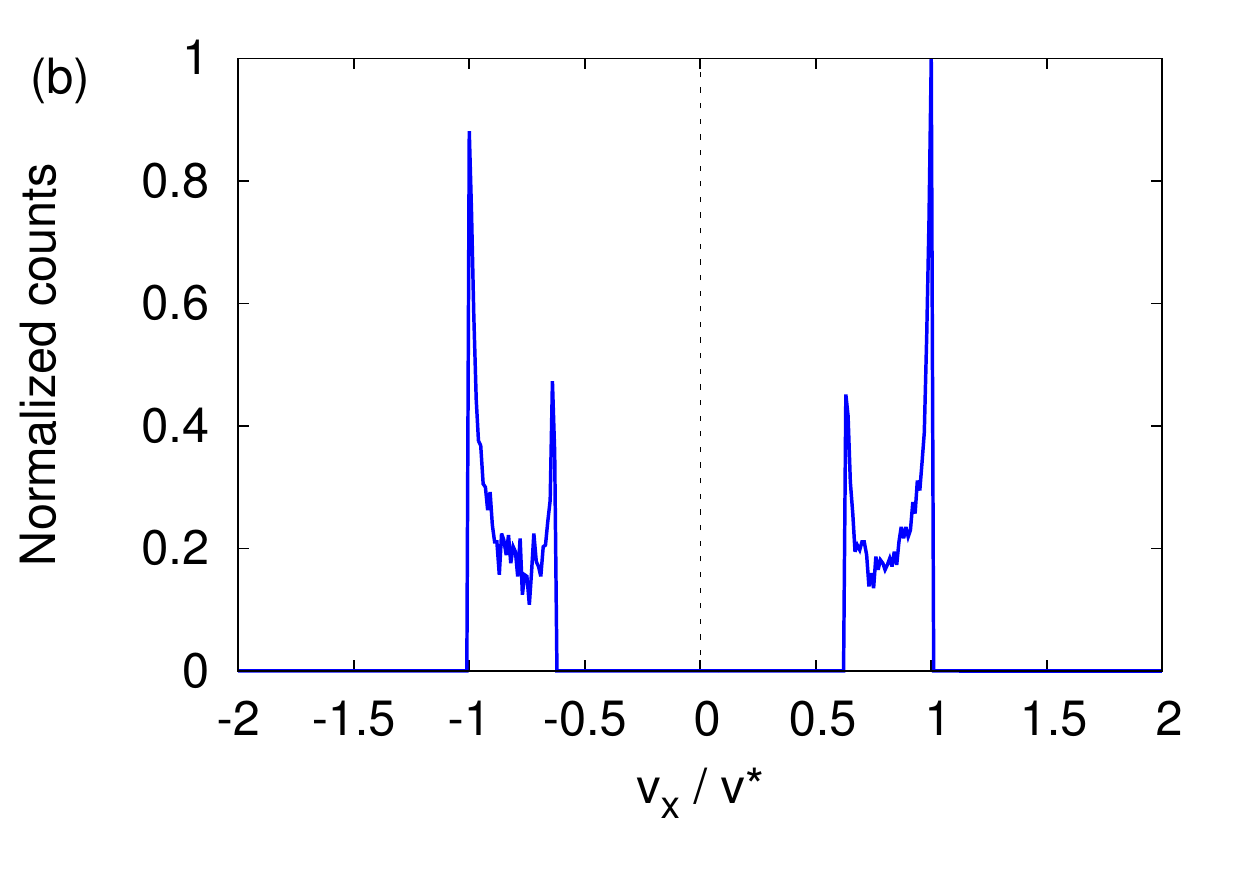}
\caption{{}
(\textbf{a}) Histogram of the transverse velocity distribution obtained by
solving the classical equations of motion Equation~(\ref{NEWT3}) for $B_0=0$,
with the initial magnetic moments distributed randomly (see text).
(\textbf{b}) Distribution of the number of particles obtained by integrating
the histogram shown in (\textbf{a}) for $v_z\in[-v^\ast,v^\ast]/100$,
{\color{black}as indicated by the gray dashed line in (\textbf{a}).}
\label{NSIM1}
}
\end{figure}

Figure~\ref{NSIM1} shows data for the case $B_0=0$.
The transverse velocity distribution looks very similar to the one shown in Figure~\ref{NSIM0}f.
In the neutron experiment~\cite{HAME75}, the neutrons that have passed through
the SG magnet are selected by means of a narrow window that moves
in one direction (say the $x$ direction) only.
The recorded neutron counts, plotted as a function of $x$, show two, very well-separated
maxima (see Figure~\ref{EXPI1}).
In analogy with the experimental procedure, we compute the one-dimensional, $x$-dependent distribution
by integrating the histogram shown in Figure~\ref{NSIM1}a for $v_z\in[-v^\ast,v^\ast]/100$.
This procedure is the computational equivalent of the moving window used in the neutron experiment.
The resulting $x$-dependent distribution is displayed in Figure~\ref{NSIM1}b.
This projected transverse velocity consists of two very well-separated distributions.
Figure~\ref{NSIM1}b strongly suggests that the presence of a magnetic field gradient
causes the incident beam of particles to split into two well-defined beams.

\begin{figure}[!htp]

\includegraphics[width=0.47\hsize]{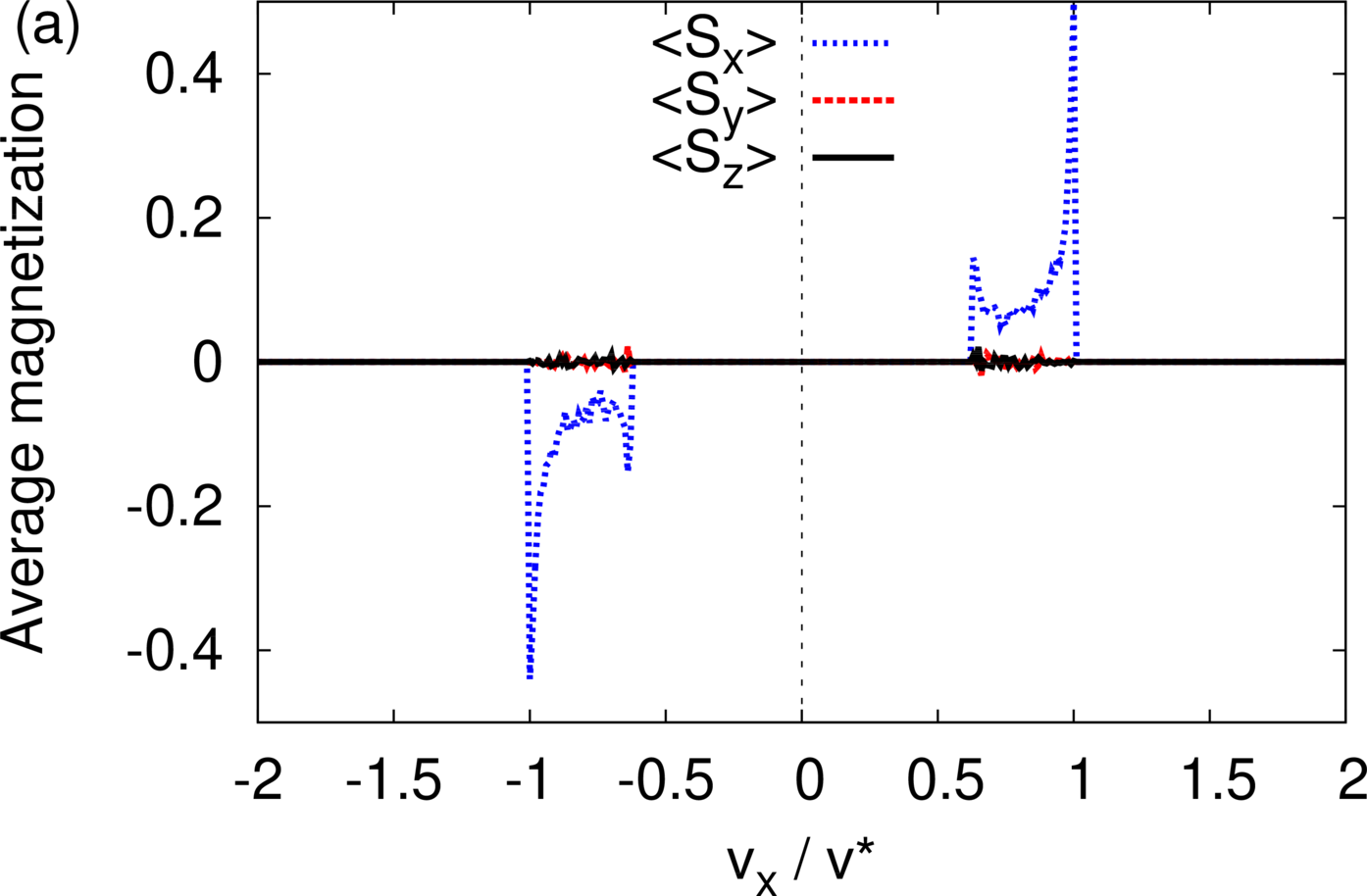}
\hspace{0.022\hsize}%
\includegraphics[width=0.47\hsize]{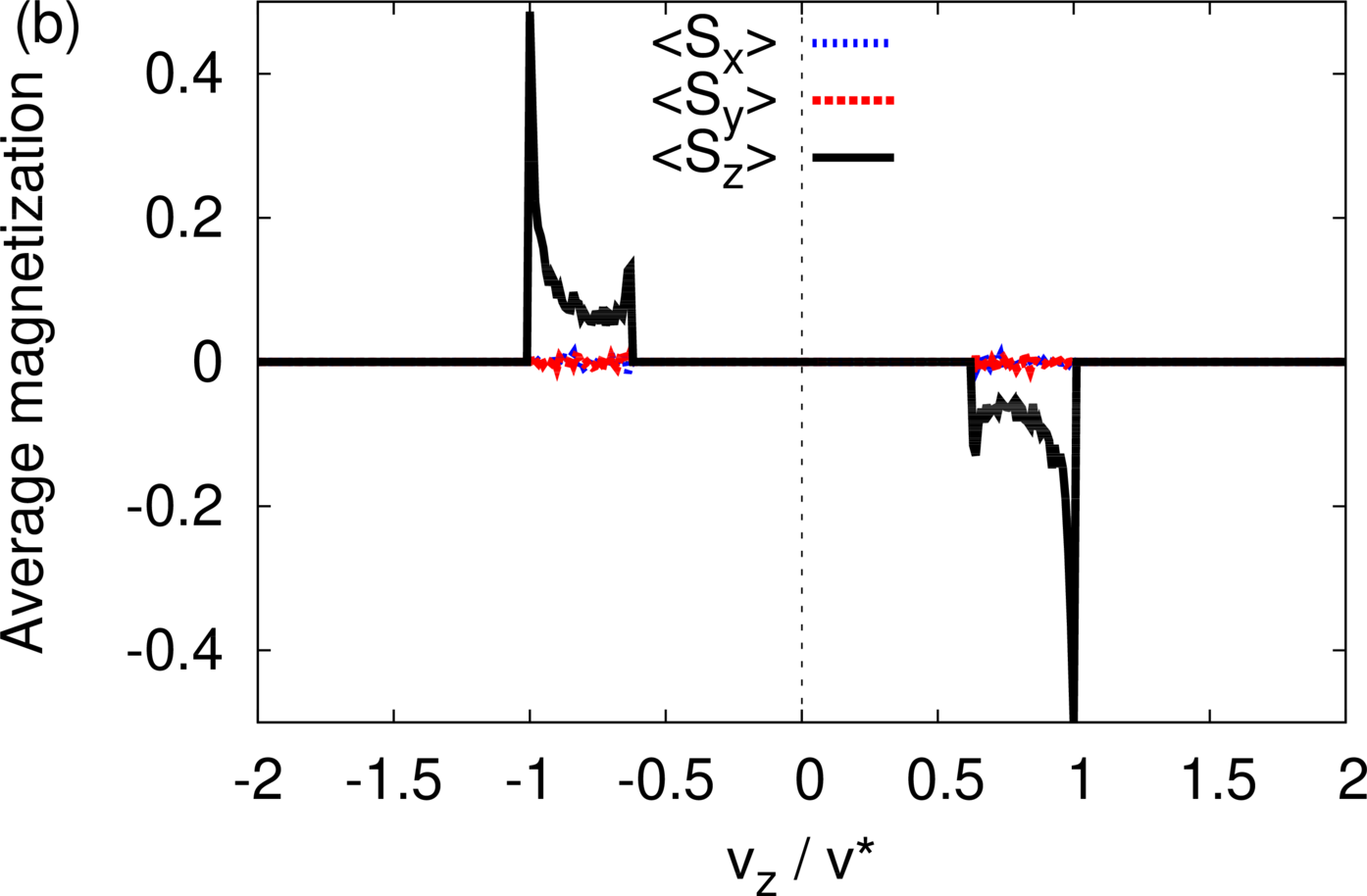}
\caption{{}
Histogram of the average of the three spin components
in the transverse velocity distribution obtained by
solving the classical equations of motion Equation~(\ref{NEWT3})
with the initial magnetic moments distributed randomly (see text).
(\textbf{a}) Average calculated by integrating the spin data for $v_z\in[-v^\ast,v^\ast]/100$.
(\textbf{b}) Average calculated by integrating the spin data for $v_x\in[-v^\ast,v^\ast]/100$.
\label{NSIM2}
}
\end{figure}

In Figure~\ref{NSIM2}, we present the corresponding data for
three spin components $S^x$, $S^y$, and $S^z$, obtained
by averaging the respective values for $v_z\in[-v^\ast,v^\ast]/100$ (Figure~\ref{NSIM2}a)
and $v_x\in[-v^\ast,v^\ast]/100$ (Figure~\ref{NSIM2}b), respectively.
Both figures clearly show that the presence of the magnet field gradient
causes the initially randomly oriented spins $\bS$
to preferably align along the direction of transverse propagation.

More specifically, focusing on the peaks at $v_z/v^\ast=\pm1$ in Figure~\ref{NSIM2}b,
we find that the particles with $S^z\approx1/2$ (and $S^x\approx0$, $S_y\approx0$)
acquired a negative transverse velocity,  whereas those with $S^z\approx-1/2$ (and $S^x\approx0$, $S^y\approx0$)
acquired a positive transverse velocity, in qualitative agreement with the quantum-theoretical description
(see Section~\ref{QSIM}).
Similarly, looking at Figure~\ref{NSIM2}a, we conclude that particles
with $S^x\approx1/2,-1/2$ (and $S^y\approx0$, $S^z\approx0$)
acquired a positive (negative) transverse velocity, also
in qualitative agreement with the quantum-theoretical description (see Section~\ref{QSIM}).
The fact that for the $x$-direction, positive and negative are interchanged with
respect to the case of the $z$-direction is a direct consequence of the
different signs of the corresponding components of the magnetic field (see Equation~(\ref{NEWT4})).

Viewed along one direction, e.g., the $z$-direction, there are two well-separated
beams, each of which has a well-defined magnetization.
Thus, in the absence of the uniform magnetic field ($B_0=0$),
the classical Newtonian model yields a one-dimensional
profile that displays all signatures of the ``quantization of the magnetic moment''.
Or, put differently, unless the uniform magnetic field $B_0$ is sufficiently strong,
the classical Newtonian model predicts ``quantization of the magnetic moment'' in any direction.

\begin{figure}[!htp]

\includegraphics[width=0.40\hsize]{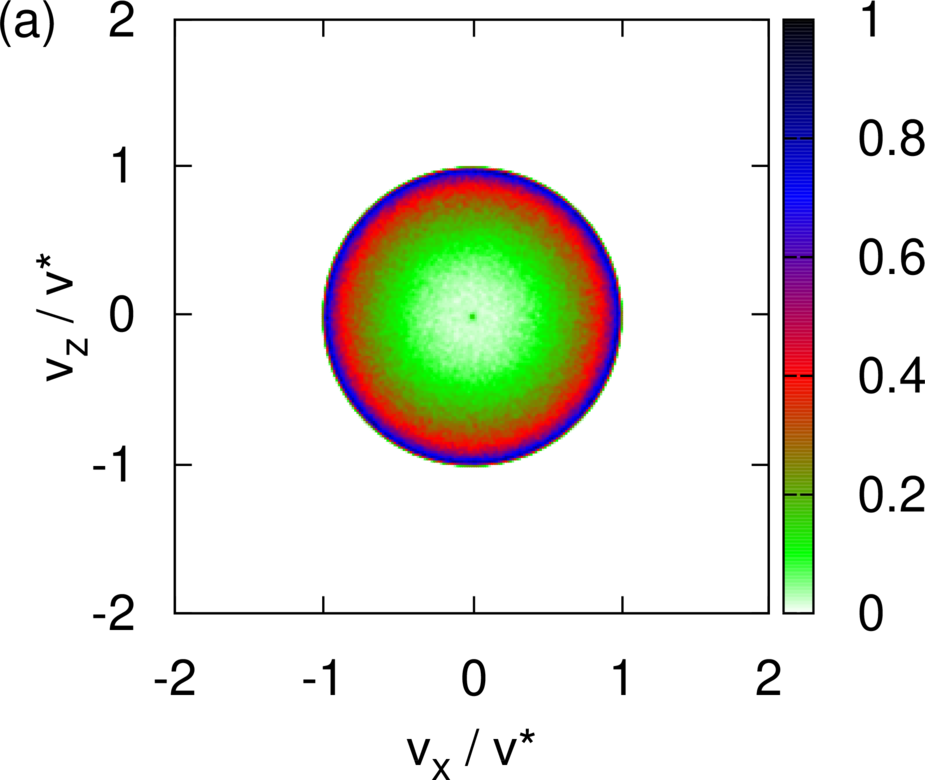}
\hspace{0.022\hsize}%
\includegraphics[width=0.55\hsize]{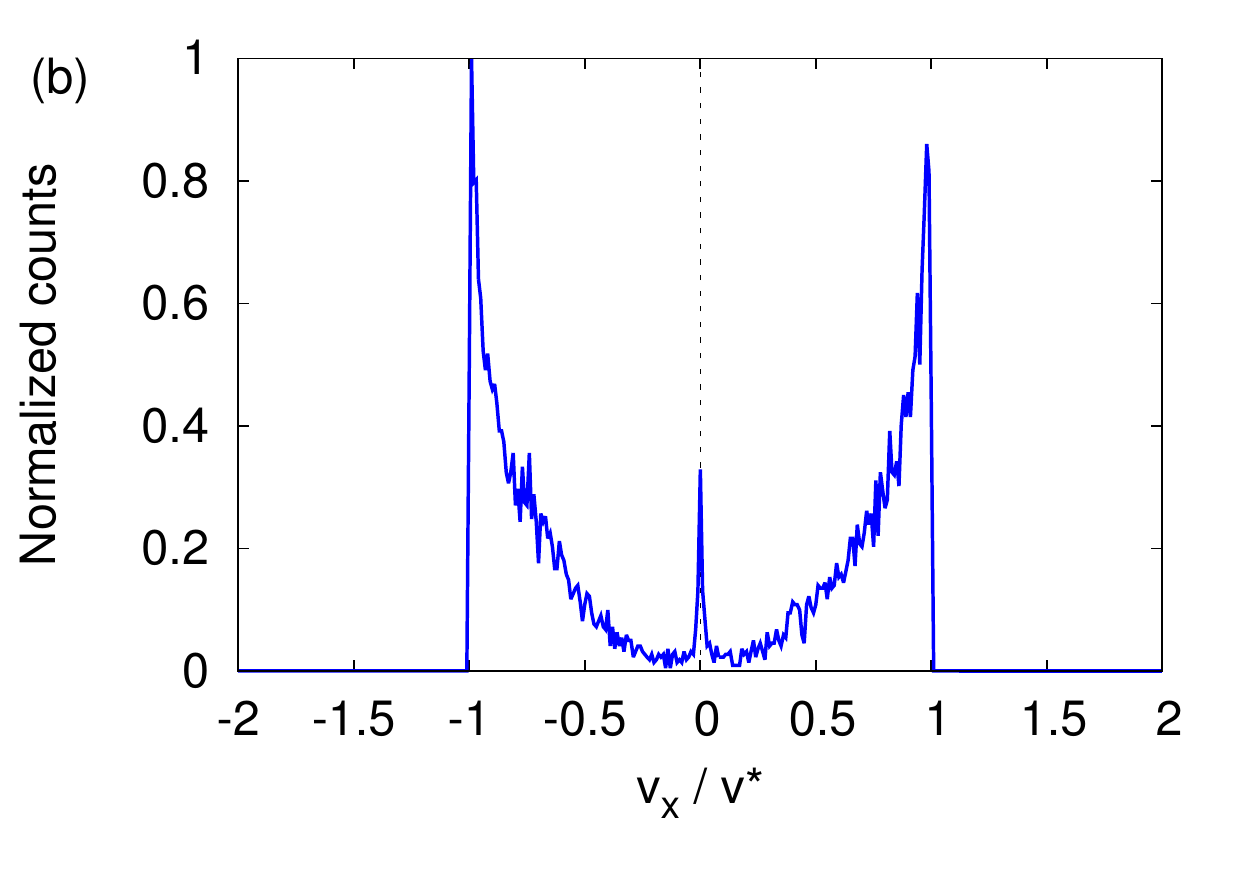}
\caption{{}
(\textbf{a}) Same as Figure~\ref{NSIM1}a except that as the particles depart from the source,
the variance of the transverse velocity
{\color{black}$\sigma_v=0.28v^\ast$.}
(\textbf{b}) Distribution of the number of particles obtained by integrating
the histogram shown in (a) for $v_z\in[-v^\ast,v^\ast]/100$.
The distribution obtained by integrating the same histogram for $v_x\in[-v^\ast,v^\ast]/100$
looks identical and is therefore not shown.
\label{NSIM4}
}
\end{figure}

For completeness, Figure~\ref{NSIM4} shows how a spread in the initial transverse velocities
affects the final transverse velocity distribution for $B_0=0$.
Clearly, the main features displayed in Figure~\ref{NSIM1} are prominently present.

\begin{figure}[!htp]

\includegraphics[width=0.40\hsize]{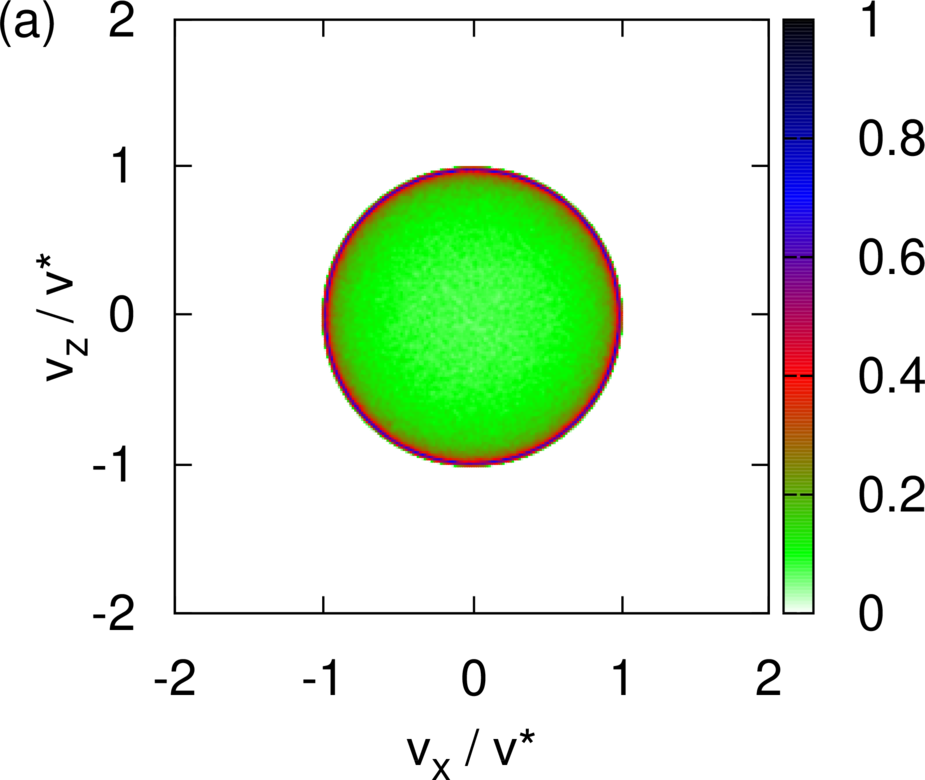}
\hspace{0.022\hsize}%
\includegraphics[width=0.55\hsize]{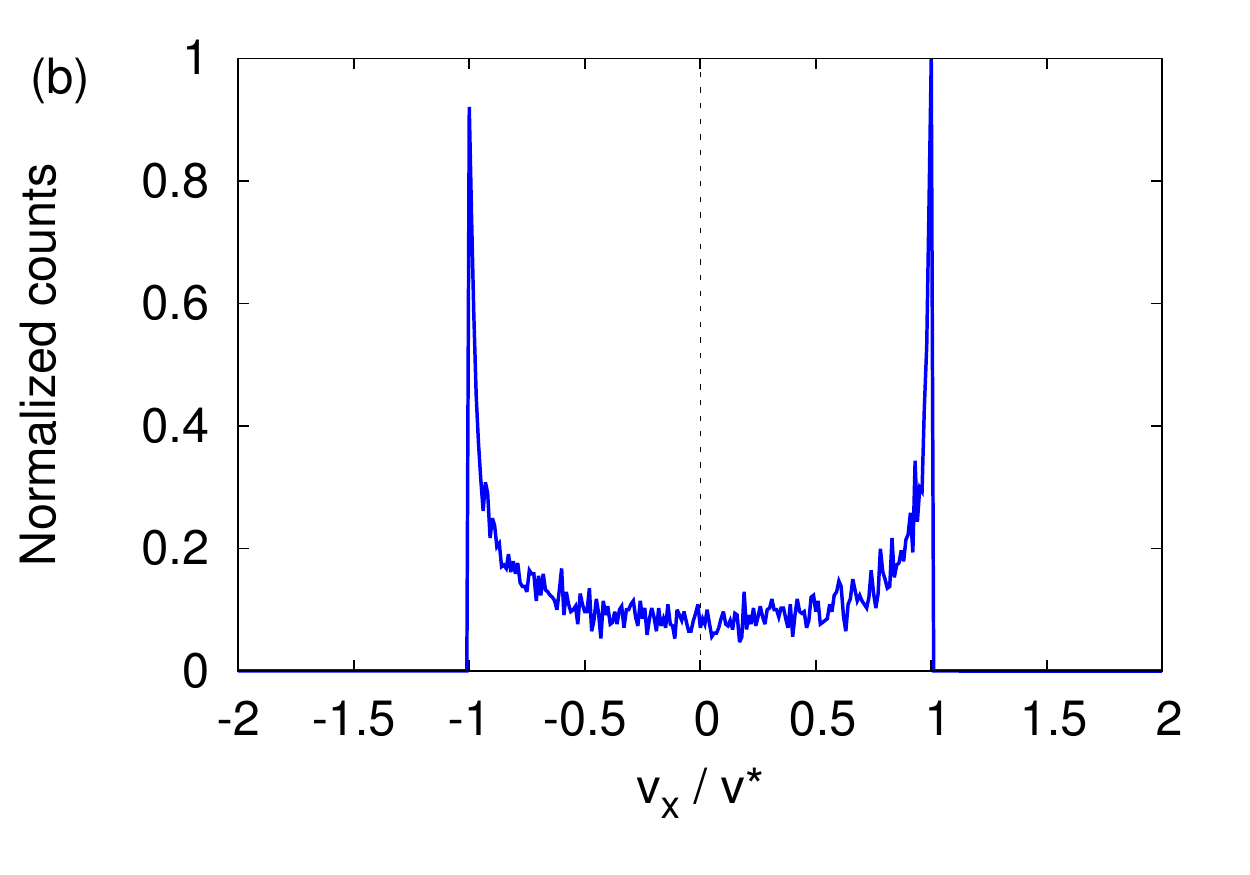}
\caption{{}
Same as Figure~\ref{NSIM1} except that the parameters
for neutrons have been replaced by the parameters
for {\color{black}imaginary silver particles}.
\label{NSIM5}
}
\end{figure}

Finally, Figure~\ref{NSIM5} shows that performing the classical
simulation using model parameters appropriate for {\color{black}imaginary silver particles}
instead of neutrons does not change the qualitative features
of the transverse velocity distribution.
Compared to neutrons (see Figure~\ref{NSIM1}b),
the main difference is that the transverse velocity distribution is more spread out
over the circle with radius $v^\ast$ (see Figure~\ref{NSIM5}b)

\section{Quantum-Theoretical Model}\label{QTM}
The Hamiltonian describing a neutral, spin-1/2 particle of mass $m$
subject to a time-independent magnetic field $\bB=\bB(\bx)$ reads
\begin{eqnarray}
H &=&\frac{1}{2m}{\bp}^2 - \frac{\hbar\gamma}{2}\bB(\bx)\cdot\bm\sigma
\;,
\label{QTM0}
\end{eqnarray}
where $\bp=(p_x,p_y,p_z)=-i\hbar\bm\nabla$ are the momentum operators,
$\bm\sigma=(\sigma^x,\sigma^y,\sigma^z)$ are the three Pauli matrices
and $\gamma$ is the gyromagnetic ratio.
The magnetic field $\bB(\bx)$ is given by Equation~(\ref{QTM1}).

From Equations~(\ref{QTM1}) and~(\ref{QTM0}), it follows immediately that $[p_y,H]=0$,
that is, the momentum in the $y$-direction is conserved.
In other words, the motion of the particles in the $y$-direction is that of a free particle; therefore, in the region where the magnetic field is nonzero,
the quantum-theoretical problem effectively amounts to solving the TDPE
\begin{eqnarray}
i\frac{\partial}{\partial t} |\Psi(t)\rangle&=&\left[
-\frac{\hbar}{2m}\left(
\frac{\partial^2}{\partial x^2}
+\frac{\partial^2}{\partial z^2}
\right)
-\frac{\gamma B_0}{2}\sigma^z
-\frac{\gamma B_1}{2} z \sigma^z
+\frac{\gamma B_1}{2} x \sigma^x
\right]
|\Psi(t)\rangle
\;,
\label{QTM2}
\end{eqnarray}
for the two-component spinor
\begin{eqnarray}
\langle x,z|\Psi(t)\rangle=
\left(\begin{array}{ll} \Psi_{{+1}}(x,z,t) \\ \Psi_{{-1}}(x,z,t) \end{array}\right)
\;,
\label{QTM3}
\end{eqnarray}
where the subscript $s=\pm1$ refers to the eigenvalues $s$ of the $\sigma^z$ operator.

In Appendix~\ref{appQTM}, we discuss the details of the analytical and numerical
tools we use to solve Equation~(\ref{QTM2}).
\subsection{Quantum Theory: Simulation Results}\label{QSIM}

\begin{figure}[!htp]

\includegraphics[width=0.47\hsize]{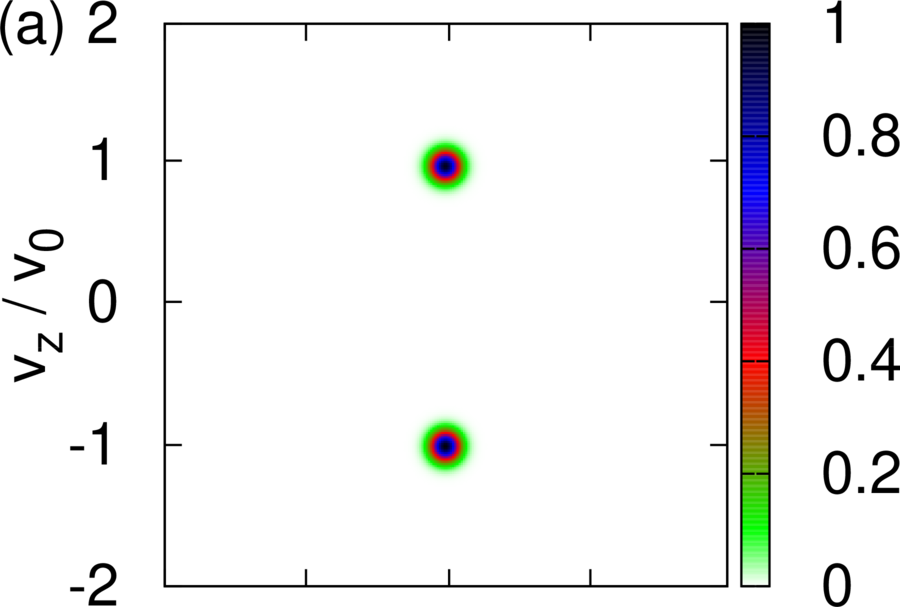}
\hspace{0.022\hsize}%
\includegraphics[width=0.47\hsize]{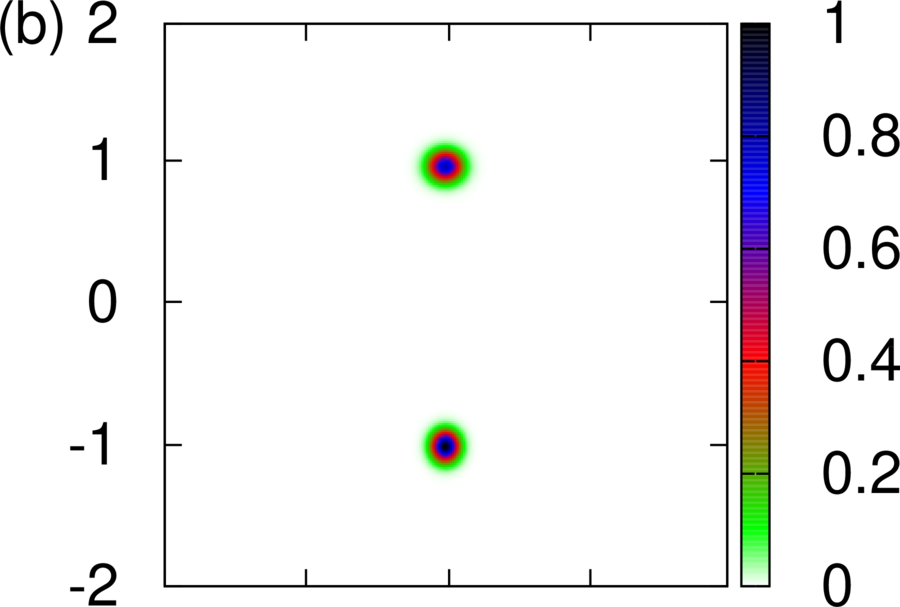}\\
\vspace{0.022\hsize}%
\includegraphics[width=0.47\hsize]{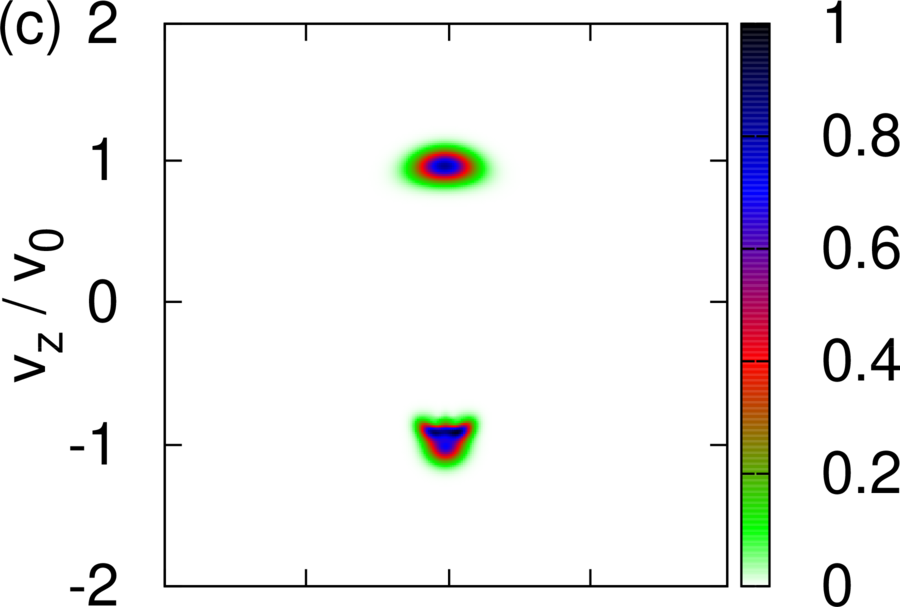}
\hspace{0.022\hsize}%
\includegraphics[width=0.47\hsize]{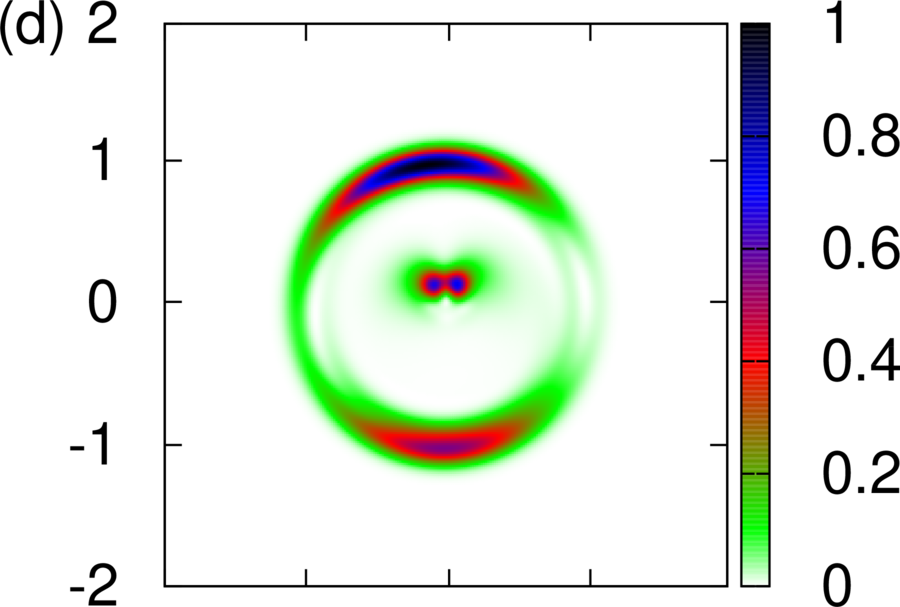}\\
\vspace{0.022\hsize}%
\includegraphics[width=0.47\hsize]{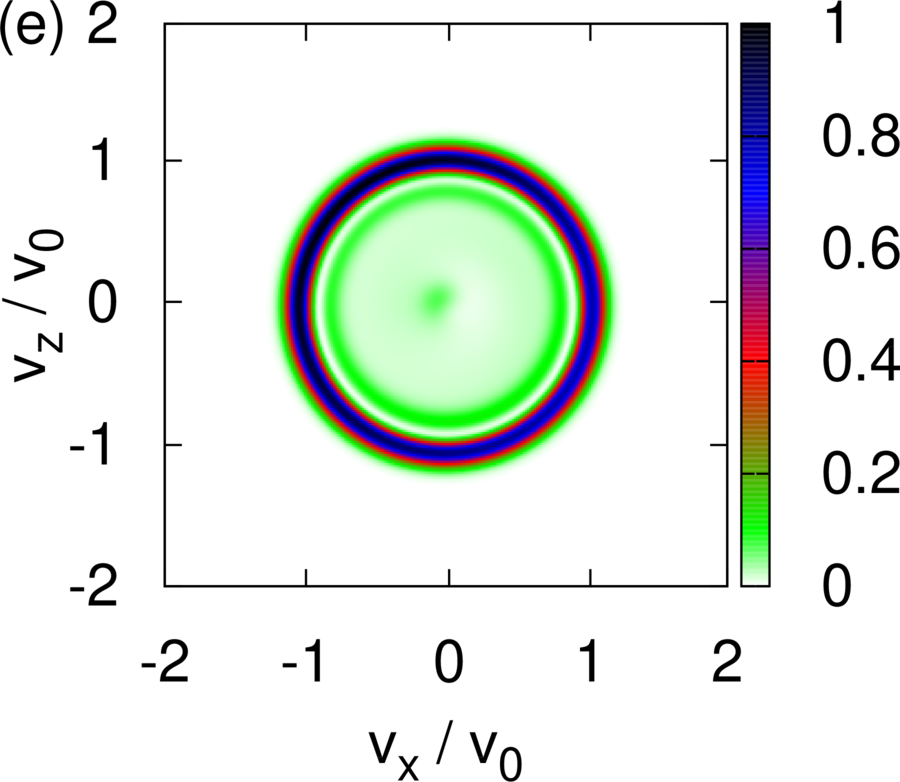}
\hspace{0.022\hsize}%
\includegraphics[width=0.47\hsize]{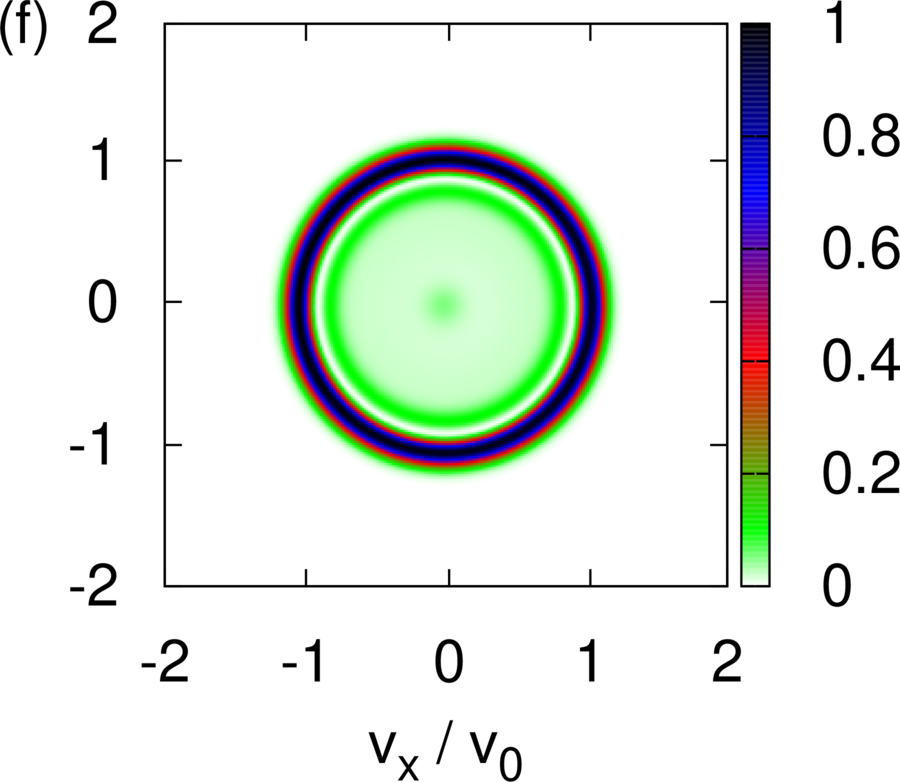}
\caption{{}
Probability distribution $|\langle v_x,v_z|\Phi(t^\ast/10)\rangle|^2$ ($v_0=v^\ast/10$)
obtained by solving the TDPE Equation~(\ref{HAMI8}) with the initial state given by Equation~(\ref{INIT0}).
Initially, the {(dimensionless) variance $\sigma=0.1$} and the spin state is $(|\uparrow\rangle + |\downarrow\rangle)/\sqrt{2}$.
(\textbf{a}) $B_0=1\; \mathrm{T}$;
(\textbf{b}) $B_0=0.1\; \mathrm{T}$;
(\textbf{c}) $B_0=0.01\; \mathrm{T}$;
(\textbf{d}) $B_0=0.001\; \mathrm{T}$;
(\textbf{e}) $B_0=0.0001\; \mathrm{T}$;
(\textbf{f}) $B_0=0.00001\; \mathrm{T}$.
\label{QSIM0}
}
\end{figure}

Figure~\ref{QSIM0} shows the transverse velocity distribution
$|\langle x,z|\Phi(t^\ast/10)\rangle|^2$
obtained by solving the TDPE Equation~(\ref{HAMI8}) for various strengths of
the uniform magnetic field and up to the time $t^\ast/10$ at
which, in the Newtonian model, the neutrons would have left the region in which
the magnetic field is present.

For a sufficiently strong uniform magnetic field, e.g., $B_0=1\,\mathrm{T}$,
the transverse velocity distribution is bimodal with well-separated maxima
at $v_z\approx\pm v_0$; see Figure~\ref{QSIM0}a.
The SG magnet then functions as an (almost perfect) filtering device,
yielding particle beams which {\sl may} be labeled by the eigenvalues of the $\sigma^z$ Pauli matrix.


We wrote {\sl may} because a meaningful assignment in terms of the eigenvalues $\sigma^z$ requires
that if we send the beam of particles through a second SG magnet
with its strong uniform magnetic field along the $z$-axis,
the particles should emerge in one and the same beam only.

More generally, if we use a filter device to label different outcomes,
subsequent repeated filtering by identical devices should leave the labeling intact~\cite{SCHW59}.
If it does not, the original assignment is useless.

Thus, to verify that an SG magnet
with its strong uniform magnetic field along the $z$-axis acts as a spin-filtering device,
we repeat the simulation with $B_0=1\; \mathrm{T}$ and initial spin state $|\uparrow\rangle$.
The resulting transverse velocity distribution is
the same as the one in Figure~\ref{QSIM0}a with the top spot removed (image not shown).
Thus, with a strong static field $B_0$, the SG magnet indeed acts as an ideal
filtering device.

In the quantum-theoretical treatment, the spin is quantized by construction; therefore, the observed splitting of the beam cannot be regarded as evidence
for the quantization of the spin; however, for large $B_0$, the quantized spin model shows
that the SG magnet splits the beam (in agreement with experiment)
whereas the Newtonian model does not (in disagreement with experiment),
exposing a fundamental shortcoming of the latter.

As in the classical case (see Figure~\ref{NSIM0}a--f), the transverse velocity distribution changes
drastically with each reduction of $B_0$ by an order of magnitude; see Figure~\ref{QSIM0}a--f.
The distributions for large (Figure~\ref{QSIM0}a,b) and small (Figure~\ref{QSIM0}e,f)
values of the uniform magnetic field $B_0$ are robust to changes of $B_0$
but for intermediate values of $B_0$ (Figure~\ref{QSIM0}c,d), it is hard
to predict the distribution.
The distributions shown in Figure~\ref{QSIM0}e,f look very similar to their
classical counterparts shown in Figure~\ref{NSIM0}e,f but differ in the details.

Maxwell's equation dictates that (with our choice of the frame of reference)
the Hamiltonian of an SG experiment should contain terms in both $\gamma\sigma^x B_1$ and
$\gamma\sigma^z B_1$, which implies that the magnetization (in any direction) is not conserved.
Therefore, unless $B_0\rightarrow\infty$, the eigenvalues of $\sigma^z$ cannot be used to label
the eigenstates of the Hamiltonian.
In other words, there are situations, choices of the model parameters, for which
the SG magnet cannot be used to define the quantization direction of the spin~\cite{POTE05,Hsu2011}.

\begin{figure}[!htp]

\includegraphics[width=0.45\hsize]{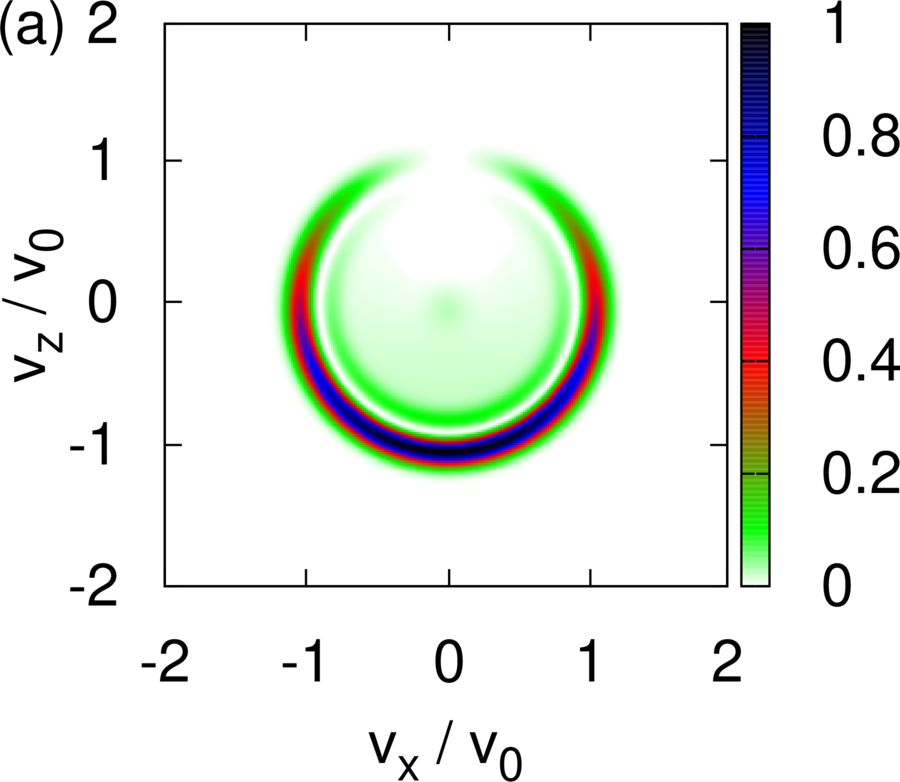}
\hspace{0.022\hsize}%
\includegraphics[width=0.45\hsize]{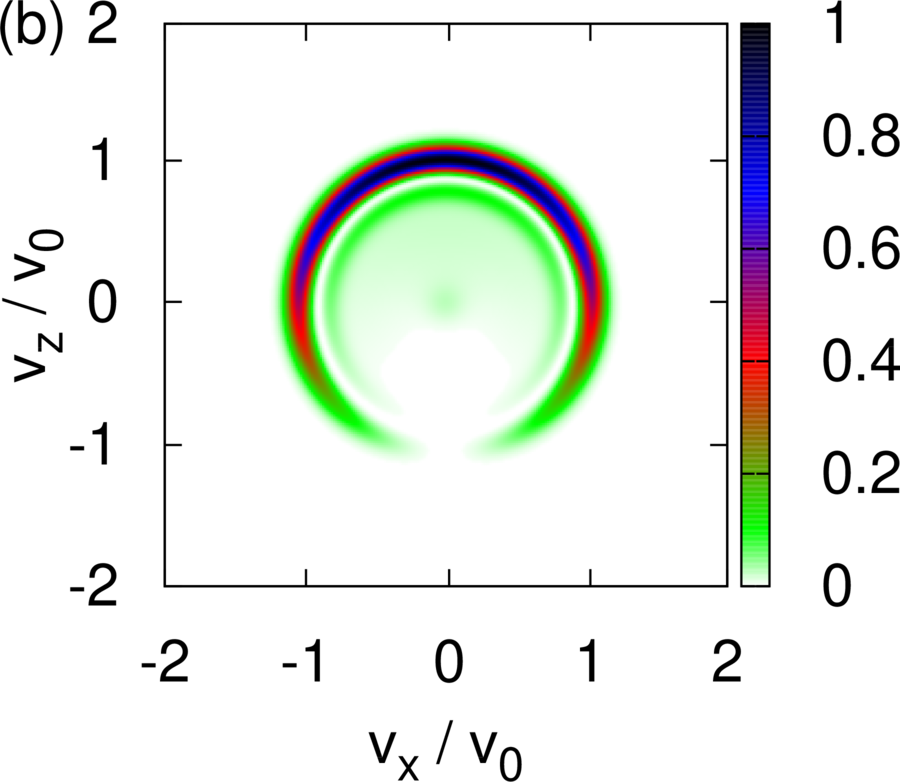}
\caption{{}
(\textbf{a}) Probability distribution $|\langle v_x,v_z|\Phi(t^\ast/10)\rangle|^2$ ($v_0=v^\ast/10$)
of the transverse velocity distribution obtained by
solving the TDPE Equation~(\ref{HAMI8}) with the initial state given
by Equation~(\ref{INIT0}) and $B_0=0$.
Initially, the {(dimensionless) variance $\sigma=0.1$} and the spin state is $|\uparrow\rangle$.
(\textbf{b}) Same as (a)
except that the initial spin state is $|\downarrow\rangle$.
\label{QSIM1}
}
\end{figure}

We study this aspect by solving the TDPE for the initial state given by Equation~(\ref{INIT0})
with $\theta=\alpha=0$, that is for the initial spin states $|\uparrow\rangle$ and $|\downarrow\rangle$
and $B_0=0$.
The transverse velocity distributions are shown in Figure~\ref{QSIM1}a,b.
If the initial spin state is $|\uparrow\rangle$ ($|\downarrow\rangle$), the wave packet dominantly propagates
along the $-z$-direction and $+z$-direction; see Figure~\ref{QSIM1}a,b, respectively.
The term $\gamma\sigma^x B_1$ causes both components of the wave function to spread in all directions,
producing the sickle-like shapes in Figure~\ref{QSIM1}a,b.
Not surprisingly, the sum of Figure~\ref{QSIM1}a,b yields an image that looks very much like
Figure~\ref{QSIM0}f.

\begin{figure}[!htp]

\includegraphics[width=0.49\hsize]{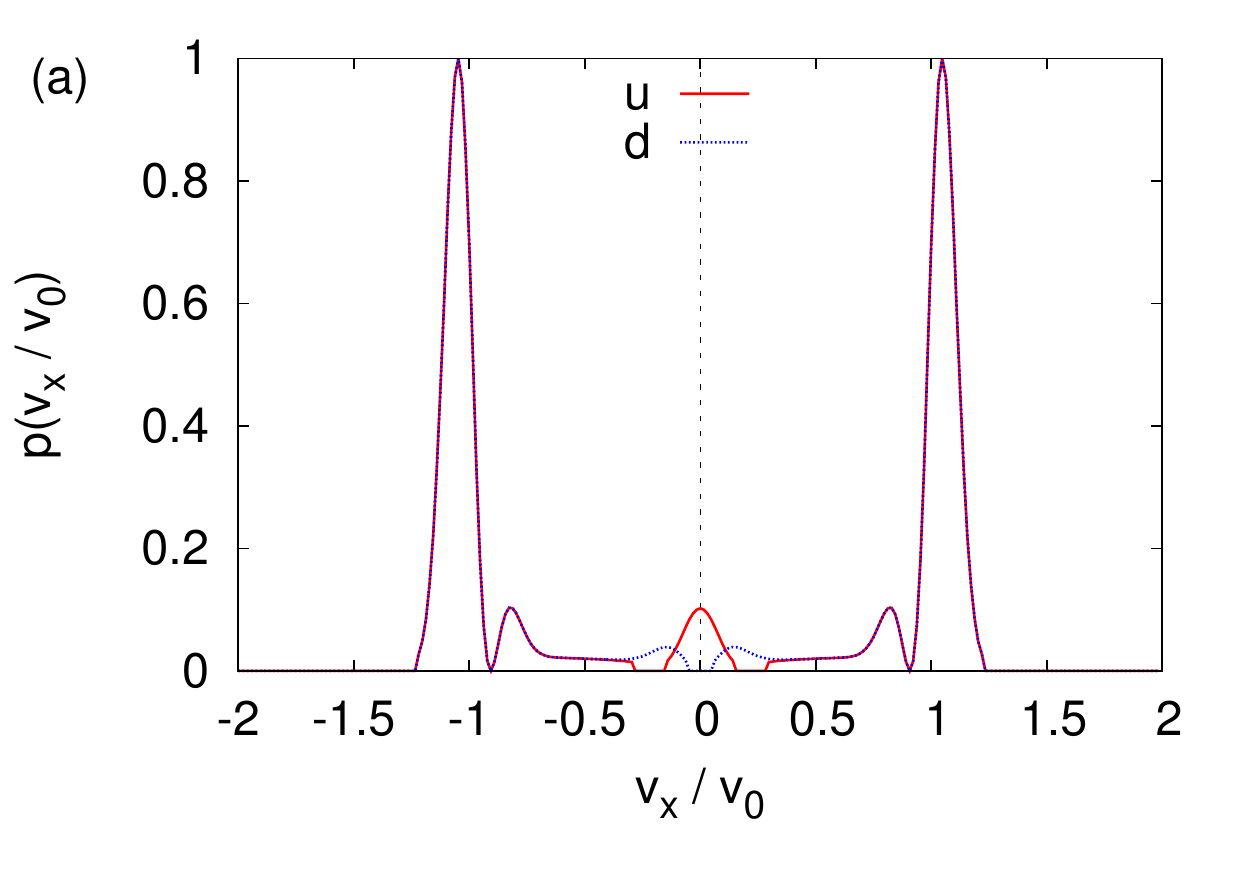}
\includegraphics[width=0.49\hsize]{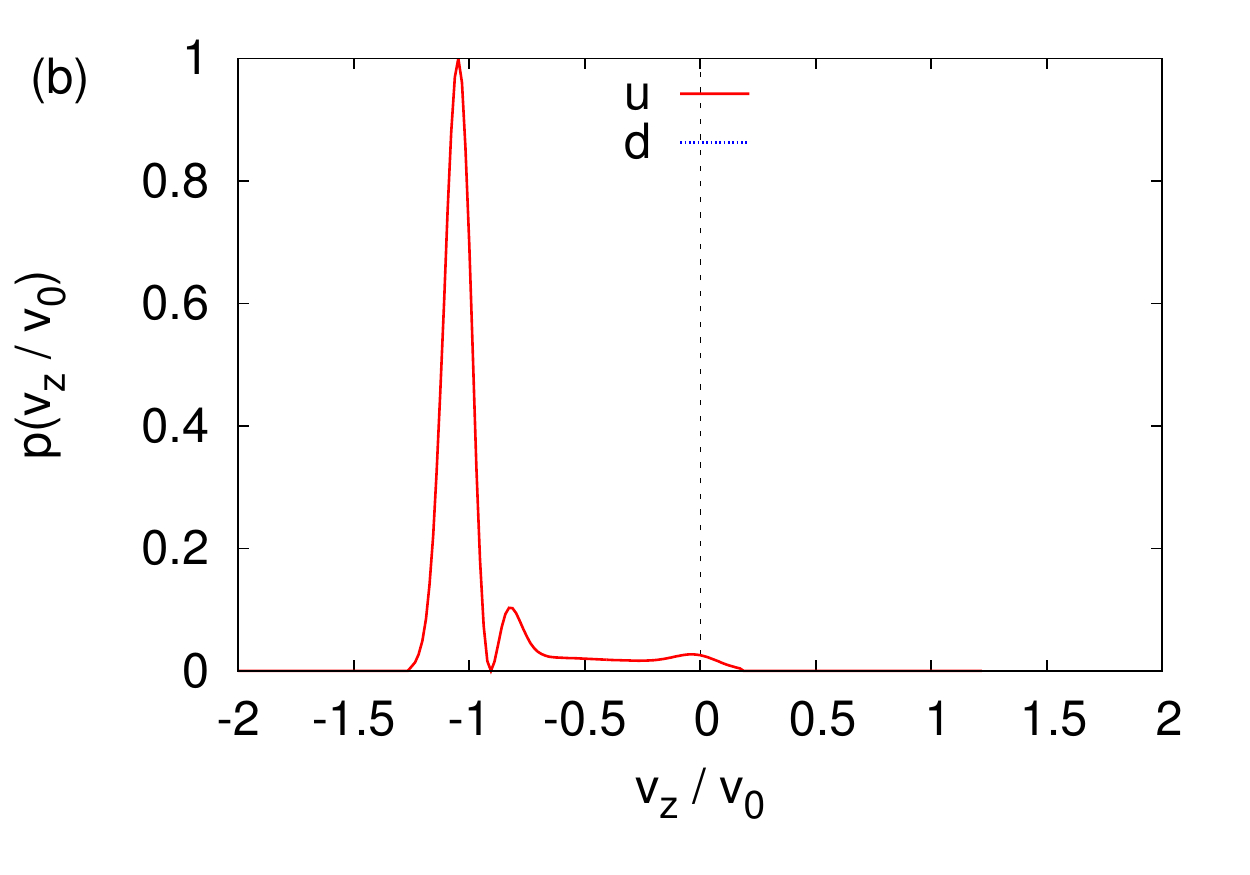}
\caption{{}
(\textbf{a}) One-dimensional probability distributions 
{\color{black}
$p(v_x)=|\Phi_{+1}(v_x,v_z=0,t^\ast/10)|^2$ (u, solid line) and
$p(v_x)=|\Phi_{-1}(v_x,v_z=0,t^\ast/10)|^2$ (d, dotted line),
extracted from the data shown in Figure~\ref{QSIM1}a.
Except for {$|v_x/v_0|\le0.2$}, the difference between two distributions is too small to be visible in the plot.
(\textbf{b}) One-dimensional probability distributions 
$p(v_z)=|\Phi_{+1}(v_x=0,v_z,t^\ast/10)|^2$ (u, solid line) and
$p(v_z)=|\Phi_{-1}(v_x=0,v_z,t^\ast/10)|^2$} (d, dotted line),
extracted from the data shown in Figure~\ref{QSIM1}a.
The probability distributions $|\Phi_{-1}(v_x=0,v_z,t^\ast/10)|^2$ is too small
to be visible in the plot.
Except for {$|v_x/v_0|\le0.2$}, the difference between two distributions is too small to be visible in the plot.
For presentation purposes, each distribution is normalized such that its maximum is one.
As in Figure~\ref{QSIM1}, $v_0=v^\ast/10$.
\label{QSIM2}
}
\end{figure}

Figure~\ref{QSIM2}a,b shows the corresponding probability distributions for
the $|\uparrow\rangle$ and $|\downarrow\rangle$ components of the wave function,
projected onto the $z=0$ and $x=0$ axis, respectively.

If, in an experiment such as the one with cold neutrons~\cite{HAME75},
one would only count particles by moving a narrow window along the $x$-direction,
the distribution shown in Figure~\ref{QSIM2}a would lead us to conclude
that the SG magnet has split the beam into parts.
On the other hand, measuring with a moving window along the $z$-direction yields
the distribution shown in Figure~\ref{QSIM2}b, which forces us to conclude that
only the $|\downarrow\rangle$ component is present in the outgoing beam.
Indeed, the intensity of the $|\uparrow\rangle$ component is several
orders of magnitude smaller than the one of the $|\downarrow\rangle$ component.
{\color{black}For $B_0\gtrsim0$, the SG magnet does not act as a spin filter.}

It may be of interest to note that if an SG magnet is used
to measure the magnetic moment to, e.g., atomic clusters~\cite{BACH18},
the value of $B_0$ does not matter much.
The positions of the peaks in the one-dimensional
distributions, which are the same for large and zero uniform magnetic field $B_0$,
suffice to determine the value of the magnetic moment.

\subsection{Quantum Theory: Simplified Model}\label{SIMPLE}

The TDPE Equation~(\ref{QTM2}), with the term in $\sigma^x$ removed, is an excellent
approximation to the full TDPE Equation~(\ref{QTM2})
if the uniform magnetic field is strong enough, e.g., if $B_0=1\,\mathrm{T}$.
Then, we may also replace $\sigma^z$ by $\bm\sigma\cdot\bb$
where $\bb=\bB/\Vert \bB\Vert$ is the unit vector parallel to the strong, uniform magnetic field,
because (i) the eigenvalues of $\bm\sigma\cdot\bb$ are the same as those of $\sigma^z$ and
(ii) only the eigenvalues enter in the spin part of the simplified TDPE.
By introducing $\bb$, the latter can describe situations in which the
strong uniform magnetic field can take any orientation, as long as it is approximately perpendicular to
the $y$-direction (otherwise the argument to remove the $\sigma^x$ term may break down).

We can now simplify the description further.
Because of the one-to-one correspondence between the eigenvalue of $\bm\sigma\cdot\bb$
and the change in the transverse velocity of the outgoing particles,
we may dispose of the description of the translational degrees of freedom entirely
and represent the operation of the SG apparatus by the projection operator~\cite{BALL03}
\begin{eqnarray}
M(\bb)&=&\frac{1+\bm\sigma\cdot\bb}{2}
\;,
\label{SIMPLE0}
\end{eqnarray}
acting on the spin state $|\psi\rangle=a_\uparrow|\uparrow\rangle+a_\downarrow|\downarrow\rangle$ only.
The probability to observe a particle in the beam labeled by $s_\bb=\pm1$, one of the two eigenvalues
of $\bb\cdot\bm\sigma$, is given by
\begin{eqnarray}
P(s_\bb|\xi)&=&\langle\psi|M(\bb)|\psi\rangle
=\frac{1+s_\bb\,\cos\xi}{2}=
\left\{\begin{array}{ccc}
\cos^2\frac{\xi}{2}&,&s_\bb=+1\\
\\
\sin^2\frac{\xi}{2}&,&s_\bb=-1\\
\end{array}
\right.
\;,
\label{SIMPLE1}
\end{eqnarray}
where $\cos\xi=\bs\cdot\bb$ and $\bs=\langle\psi|\bm\sigma|\psi\rangle$.
The last expression in Equation~(\ref{SIMPLE1}) is reminiscent of Malus' law for the intensity of polarized light
passing through a polarizer.

The projector equation (Equation~(\ref{SIMPLE0})) and the probability equation (Equation~(\ref{SIMPLE1})) describe
the operation of the SG apparatus in terms of the spin-degree of freedom only.
This simplified model is often used in textbooks to elucidate
quantum measurement theory~\cite{FEYN65,BAYM74, BALL03}.
We stress that Equation~(\ref{SIMPLE1}) does not apply to the case of a weak uniform magnetic field.

Solving the TDPE Equation~(\ref{QTM2}) for $B_0=1\,\mathrm{T}$
and for the initial states Equation~(\ref{INIT0}) with
$\theta=0,\pi/6,\pi/4,\pi/3$ and $\alpha=\theta/2$
yields the expected bimodal shape of the transverse velocity distributions (data not shown).
The total probabilities for $v_z<0$ and $v_z>0$ are in excellent agreement with
the prediction based on Equation~(\ref{SIMPLE1}).

\section{Event-by-Event Simulation}\label{EVENT}

From the comparison of
Figure~\ref{NSIM0}a with Figure~\ref{QSIM0}a and also of Figure~\ref{NSIMAG0}a with \mbox{Figure~\ref{QSIMAG0}b},
it is immediately clear that the transverse velocity distributions are very different if $B_0=1\,\mathrm{T}$.
For $B_0=0$, there is no qualitative difference between the Newtonian and quantum-theoretical results.

The qualitative difference between the Newtonian and quantum-theoretical
prediction in the case of a large uniform magnetic field has been
decisive to eliminate the former as a description of the
experimental observations~\cite{STER22,Frisch1933,Gerlach1924}; however, that does not imply that quantum theory is the only viable description
of experiments in which the frequency distribution of detection events is built up one-by-one,
such as in the SG experiment.

From this broader perspective, the fundamental question to be answered is ``is it possible
to construct a process that generates event-by-event and without using knowledge about
the final distribution of events, frequency distributions that are commonly thought to
be a signature of wave interference, two-particle entanglement, uncertainty, etc.''
This question is answered in the affirmative by the event-by-event simulation approach developed in
Refs.~\onlinecite{RAED05b,ZHAO08,ZHAO08b,JIN10b,RAED12b,RAED12a,MICH14a,DONK14,RAED14a,RAED16c,WILL20b,RAED20a}.

In the case at hand, the conceptually interesting question is
whether it is possible to retain a picture of the SG experiment in which
individual particles follow trajectories while, in contrast to the Newtonian results,
the transverse velocity distribution exhibits two well-separated maxima along the line defined
by the direction of the strong static field (the $z$ direction in our case).

Remarkably, a marginal modification of Newton's equation of motion
suffices to answer this question affirmatively.
The modification consists of replacing step
\begin{enumerate}\addtocounter{enumi}{3}
\item
{If} $y\in[y_0,y_1]$ set $\bF=\gamma\;B_1 S^z\be_z-\gamma\;B_1 S^x\be_x$, 
\end{enumerate}
in which the force $\bF$ is being calculated
(see Appendix~\ref{VECTOR} for details)
by the rules
\begin{enumerate}\addtocounter{enumi}{3}
\item
{If} $y\in[y_0,y_1]$:
the {\bf first} time that the event $\Vert\bB(\bx)\Vert>0$ occurs,
that is when the particle enters the region
where $\Vert\bB(\bx)\Vert>0$, use Equation~(\ref{SIMPLE1}) with $\bs=\bS$ to
align the vector $\bS$ along the magnetic field $s_\bb\bB(\bx)$
and compute $\bF=\gamma\;B_1 S^z\be_z-\gamma\;B_1 S^x\be_x$.
\end{enumerate}

In detail, if $r \le \bS\cdot\bB(\bx)/\Vert\bB\Vert$ set $\bS=\bB/2\Vert \bB\Vert$,
otherwise set $\bS=-\bB/2\Vert \bB\Vert$.
Here $r$ is a uniform (pseudo) random number in the range $[-1/2,1/2]$ (which changes
each time before it is used).
With the new $\bS$, compute $\bF=\gamma\;B_1 S^z\be_z-\gamma\;B_1 S^x\be_x$.
For each particle, the alignment of $\bS$ is carried out only once.


One might try to argue that because the event-by-event model makes
use of Equation~(\ref{SIMPLE1}), it implicitly ``knows'' about quantum theory; however, probabilistic laws such as Equation~(\ref{SIMPLE1}) also follow from the application
of logical-inference~\cite{RAED14b,RAED18a} to the modeling of event-based processes.
This approach yields Equation~(\ref{SIMPLE1}) directly, without any reference to quantum-theoretical concepts.

In short, the key idea of the logical inference approach is that ``good'' physics experiments
must yield reproducible frequency distributions which are robust, meaning do not change much,
if the conditions under which the data was taken changes a little~\cite{RAED14b}.
In the case at hand, the frequency distribution consists of the average numbers of $+1$ and $-1$ events
and $\xi=\arccos(2\bS\cdot\bB(\bx)/\Vert\bB\Vert)$ represents the condition~\cite{RAED14b}.
Expressing the key idea mathematically leads to the requirement that the Fisher information
\begin{eqnarray}
I_\mathrm{F}(\xi)&=&\sum_{x=\pm1}\frac{1}{p(x|\xi)}
\left(\frac{\partial p(x|\xi)}{\partial \xi}\right)^2>0
\;,
\label{LI1}
\end{eqnarray}
for the probability $p(x|\xi)$ to observe the event $x=\pm1$ under the condition $\xi$
must be independent of $\xi$ and minimal~\cite{RAED14b}.
After some elementary algebra, we find that the solution of this optimization problem reads~\cite{RAED14b}
\begin{eqnarray}
p(x|\xi)&=&\frac{1\pm x\cos\xi}{2}
\;,
\label{LI2}
\end{eqnarray}
where the $\pm$ sign reflects the ambiguity in assigning $+1$ or $-1$ to one of the directions.
Quantum theory postulates Equation~(\ref{SIMPLE1}) (through the Born rule) whereas
the logical inference approach allows us to derive Equation~(\ref{SIMPLE1}) without
making reference to a concept of quantum theory.
Thus, the argument that the event-by-event algorithm implicitly refers to quantum theory does not hold.

Moreover, the modification does not change the vector character of $\bS$.
In the event-by-event model, $\bS$ can take any value on the sphere of radius 1/2,
there is no wave function, there are no Pauli spin matrices, there simply is
no element of quantum theory in the event-by-event model.

\begin{figure}[!htp]

\includegraphics[width=0.47\hsize]{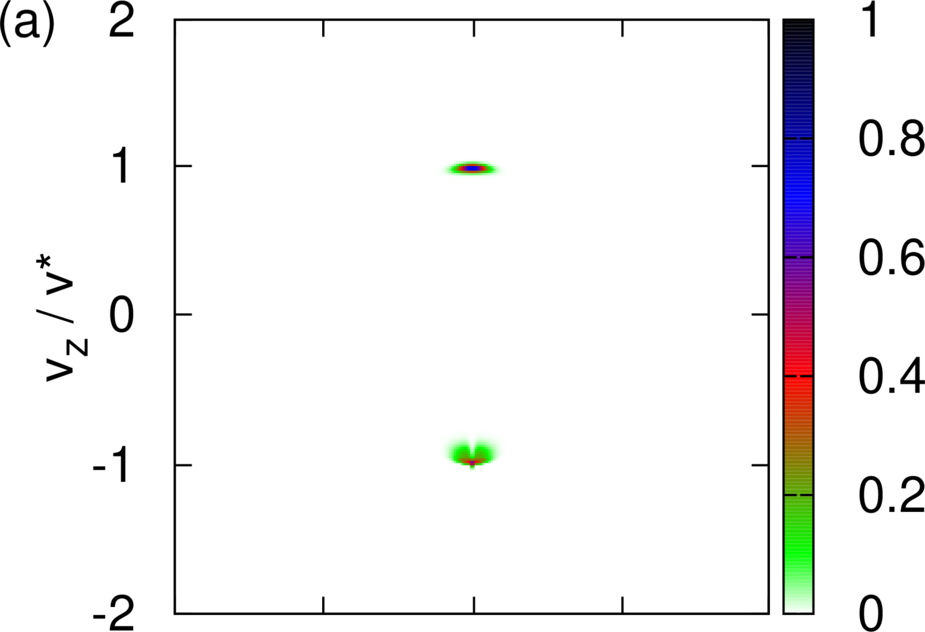}
\hspace{0.022\hsize}%
\includegraphics[width=0.47\hsize]{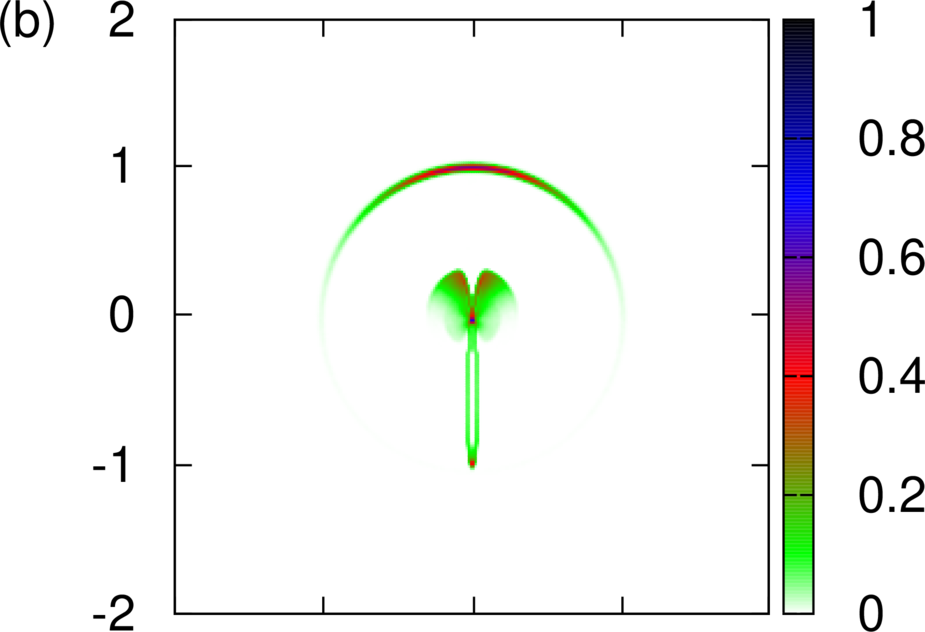}\\
\vspace{0.022\hsize}%
\includegraphics[width=0.47\hsize]{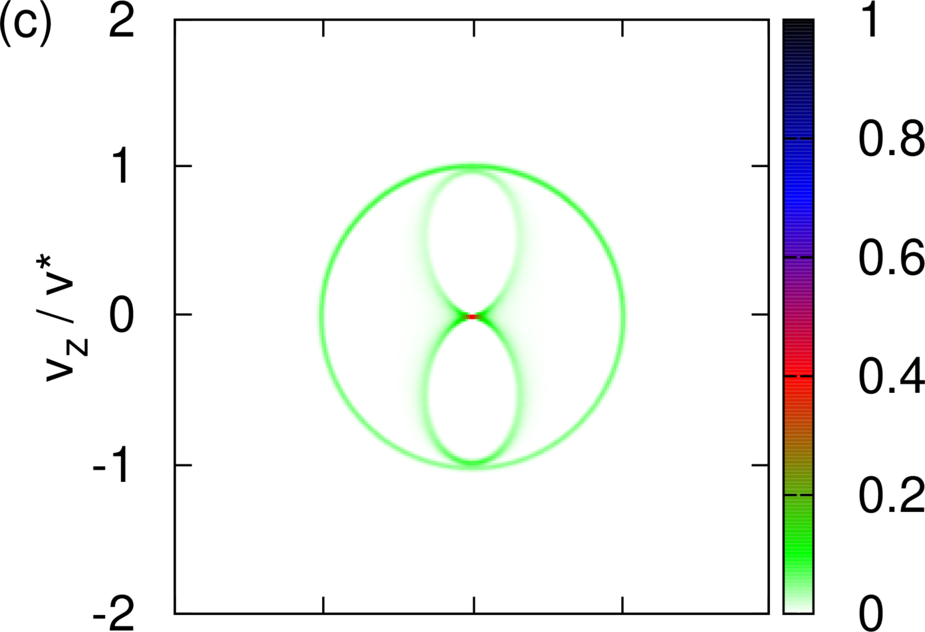}
\hspace{0.022\hsize}%
\includegraphics[width=0.47\hsize]{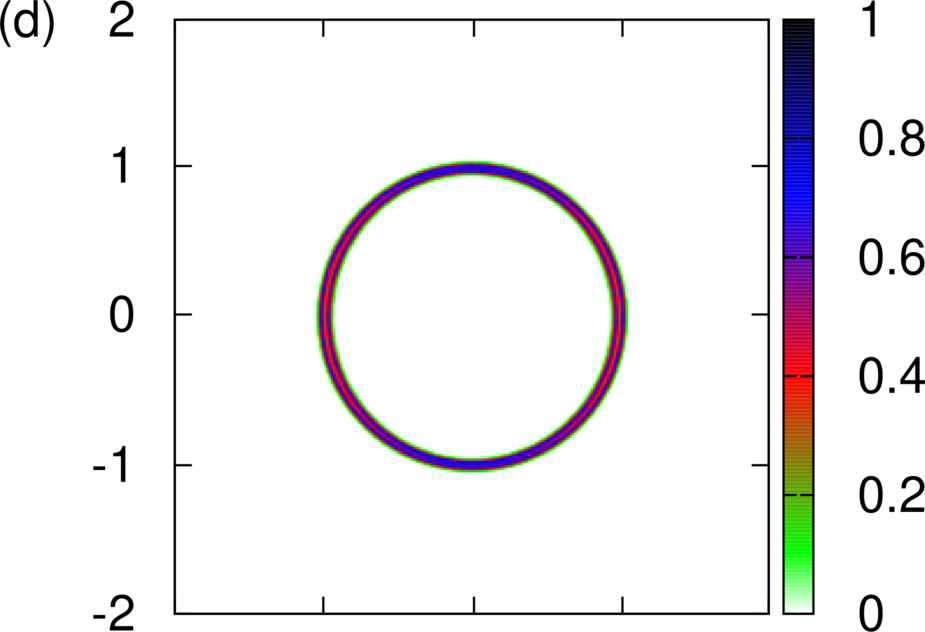}\\
\vspace{0.022\hsize}%
\includegraphics[width=0.47\hsize]{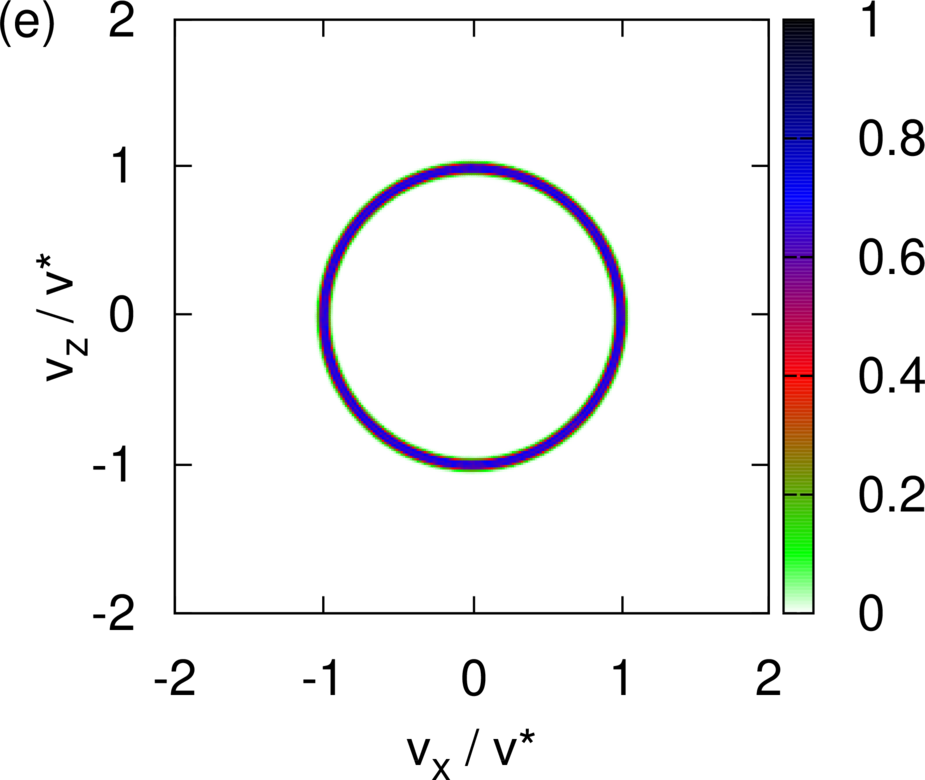}
\hspace{0.022\hsize}%
\includegraphics[width=0.47\hsize]{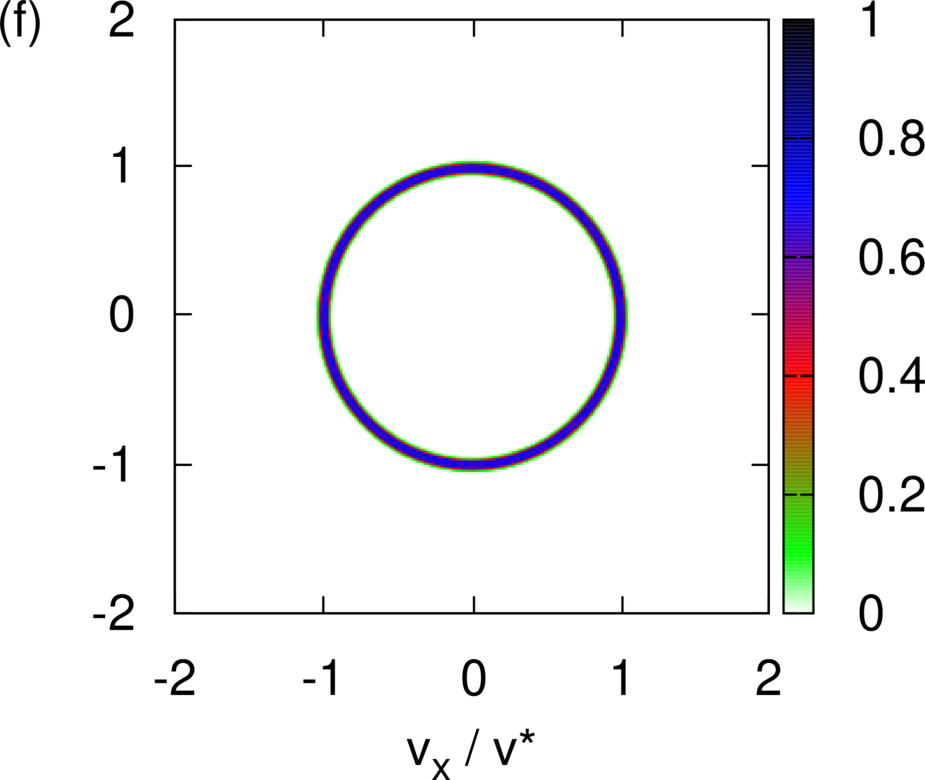}
\caption{{}
Histograms of the transverse velocity distribution obtained by event-by-event simulation.
The classical equations of motion (Equation~(\ref{NEWT3}))
are modified to include a one-time projection of the spin vector $\bS$ along the direction of the magnetic field
{\color{black}using the procedure described in the text.}
The initial magnetic moments of the particles are distributed randomly.
The variance of the transverse velocity {\color{black}$\sigma_v=0.014 v^\ast$}.
(\textbf{a}) $B_0=1\; \mathrm{T}$;
(\textbf{b}) $B_0=0.1\; \mathrm{T}$;
(\textbf{c}) $B_0=0.01\; \mathrm{T}$;
(\textbf{d}) $B_0=0.001\; \mathrm{T}$;
(\textbf{e}) $B_0=0.0001\; \mathrm{T}$;
(\textbf{f}) $B_0=0.00001\; \mathrm{T}$.
\label{EVENT0}
}
\end{figure}

Figure~\ref{EVENT0}a demonstrates that the event-based model produces
a bimodal transverse velocity distribution, in qualitative agreement
with the solution of the TDPE Equation~(\ref{QTM2}).
Clearly, the minor modification to Newton's equation has a tremendous impact
on the trajectories of the particles.
For $B_0\approx0$, the event-by-event simulations yields
the circular distribution; see Figure~\ref{EVENT0}e,f, in qualitative
agreement with both the Newtonian and quantum-theoretical description.

As a further check, we perform event-by-event simulations for $B_0=1\,\mathrm{T}$
and take as initial spin vector
$\bS=(\cos\phi\sin\theta,\sin\phi\sin\theta,\cos\theta)^{\mathrm{T}}/2$
for $\theta=0,\pi/6,\pi/4,\pi/3$, $\pi/2$, $2\pi/3$, $3\pi/4$, $5\pi/6$, $\pi$ and $\phi$
uniformly random from the interval $[0,2\pi]$ (data not shown).
The total probabilities for $v_z<0$ and $v_z>0$ are in excellent agreement with Equation~(\ref{SIMPLE1}),
that is with quantum theory.

\section{Conclusions}\label{CONC}

In all our simulations, the strength of the magnetic field gradient was fixed
and tuned to the case of an SG experiment with cold neutrons~\cite{HAME75}
while the strength of the uniform component of the magnetic field was varied.
The simulation data for {\color{black}imaginary silver particles} instead of neutrons show the same qualitative features.

In Table~\ref{tab1}, we collect the most essential features of the results
for the transverse velocity distribution obtained
from the simulation of three different models of the SG experiment.
Thereby, we have omitted many of the computational details mentioned earlier
and limit the discussion to the two extreme cases of a strong and zero uniform magnetic field.

\begin{table}[!htp]
\caption{%
Overview of the shapes of the transverse velocity distributions
obtained from by computer simulation of
three different descriptions of the SG experiment with cold neutrons.
}
\begin{ruledtabular}
\begin{tabular}{ccc}
& $B_0=0$ &  $B_0=1\,\mathrm{T}$ \\
\hline\noalign{\smallskip}
Experiment      &  ???                            &  two spots (Fig.~\ref{EXPI1})\\
Newton          &  circular (Fig.~\ref{NSIM0}(f)) &  one stripe (Fig.~\ref{NSIM0}(a))   \\
Quantum theory  &  circular (Fig.~\ref{QSIM0}(f)) &  two spots (Fig.~\ref{QSIM0}(a))\\
Event-by-event  &  circular (Fig.~\ref{EVENT0}(f))&  two spots (Fig.~\ref{EVENT0}(a))\\
\end{tabular}
\end{ruledtabular}
\label{tab1}
\end{table}


%

The first three rows of the last column in Table~\ref{tab1}
express what is known since the original SG experiment
was performed, namely that Newtonian mechanics cannot explain
the observed splitting of the particle beam~\cite{STER22,GERL24,FEYN65,BOHM51,BAYM74,BALL03}
if the uniform magnetic field component is sufficiently large.
It is exactly under this last condition that the quantum-theoretical textbook model
provides an accurate description of the time evolution of the probability distribution
while the Newtonian model does not.

However, we have also shown that a minor modification to Newton's equations of motion
yields results that are in line with the experimental observation and quantum theory.
In this event-by-event simulation approach,
the spin is described in terms of a three-dimensional vector,
not in terms of Pauli matrices.

If the strength of the uniform magnetic field $B_0$ gradually decreases,
then, for any of the three models, the changes in the transverse velocity distribution
become hard to predict analytically, unless the effect of $B_0$ becomes negligible.
Indeed, for $B_0=0$ a symmetry argument can be used to understand
why the calculated transverse velocity distribution shows a circular structure,
see column two of Table~\ref{tab1}.

However, also the case $B_0\gtrsim0$ poses some interesting interpretational issues,
depending on how the distribution of particles is measured.
If, as in the neutron experiment~\cite{HAME75}, one only records the distribution
along a particular direction,  the Newtonian model also yields a bimodal distribution.
Without additional data, the bimodality would (erroneously, see Section~\ref{QSIM}) imply that
the two beams can be labeled by the spin quantum number.

{\color{black}
From a general perspective, quantization (not to be confused with results from
quantum theory) is the process of classifying empirical data into groups and
attaching discrete labels to these groups. As mentioned at the end of Section~\ref{EXPI},
in the specific case of the neutron experiment it is clear that quantization is
the result of classification, putting data points in two groups; see Figure~\ref{EXPI1}. Once
this ``operation'' has been carried out, the compressed, new data are ``quantized''.
In our view, quantum theory provides a {powerful} mathematical framework to describe
such ``quantized data''. Within quantum theory, the spin is quantized by
definition/construction. If the ``quantized'' form of the empirical data is
described well in terms of a quantum spin model, then that is a great
achievement; however, this success does not necessarily justify the conclusion
that ``quantization'' is a property/attribute of the phenomenon that gave rise to
the empirical data. In our view, drawing this conclusion mixes up the phenomenon
that gave rise to the empirical data with a quantum model of it.
}

On the basis of SG experiments that have been performed to date,
it is not possible to distinguish between the quantum-theoretical and event-by-event model.
New, high-precision experiments are needed to rule out the latter and to allow
for a quantitative comparison between experimental and simulation data.

Furthermore, it would be of interest to perform an SG experiment in which
the uniform magnetic field is weak enough to render the description textbook model invalid.
For instance, an experiment with neutrons passing through a quadrupole magnet with a large field gradient
would allow a direct comparison with our simulations (which, if needed, can easily be adapted to other field
configurations).

\section*{Acknowledgements}
We are grateful to Bart De Raedt for critical reading of the manuscript and for making pertinent comments.
The authors gratefully acknowledge the Gauss Centre for Supercomputing e.V.
(www.gauss-centre.eu) for funding this project by providing computing time on
the GCS Supercomputer JUWELS at J\"ulich Supercomputing Centre (JSC).

%
%
%
%

\appendix
\section[\appendixname~\thesection]{Large Static Field \boldmath{$B_0$}}\label{NMRa}

In the following, we assume that the particle is inside the region
where the magnetic field $\bB=(-xB_1,0,\gamma B_0+zB_1)^{\mathrm{T}}$ is present.
Written out explicitly, the torque equation for the spin reads
\begin{subequations}
\label{NMRa1}
\begin{eqnarray}
\frac{d S^x(t)}{d t}& =&-\gamma(B_0+zB_1)S^y(t)
\;,
\label{NMRa1a}
\\
\frac{d S^y(t)}{d t}& =&\gamma(B_0+zB_1)S^x(t)+\gamma x B_1 S^z(t)
\;,
\label{NMRa1b}
\\
\frac{d S^z(t)}{d t}& =&-\gamma x B_1 S^x(t)
\label{NMRa1c}
\;.
\end{eqnarray}
\end{subequations}
Next, we only consider classical particles that follow trajectories
for which $|x(t)B_1|\ll B_0$ and $|z(t)B_1|\ll B_0$.
For example, if, as in our simulations, we choose $B_0=1 \;\mathrm{T}$ and $B_1=300\; \mathrm{T}/\mathrm{m}$,
we require trajectories that satisfy $|x(t)|\ll 3\,\mathrm{mm}$ and $|z(t)|\ll 3\,\mathrm{mm}$
which, from the neutron experiment point of view is not unreasonable.
Under these conditions, we may ignore the terms in $B_1$ in Equations~(\ref{NMRa1a}) and~(\ref{NMRa1b}) and
we obtain
\begin{eqnarray}
S^x(t)& =&S^x(0)\cos \gamma B_0t - S^y(0) \sin \gamma B_0t
\;.
\label{NMRa2}
\end{eqnarray}
Substituting Equation~(\ref{NMRa2}) into the equation of motion
\begin{eqnarray}
\frac{d v_x(t)}{d t}& =&-\frac{\hbar\gamma B_1}{m}S^x(t)
\;,
\label{NMRa3}
\end{eqnarray}
and integrating over time gives
\begin{equation}
v_x(t)=v_x(0)+\beta S^x(0)\sin\gamma B_0 t - \beta S^y(0)(1-\cos\gamma B_0 t)
\;,
\label{NMRa4}
\end{equation}
where $\beta=\hbar B_1/{mB_0}$.
Integrating Equation~(\ref{NMRa4}) over time once more results in
\begin{eqnarray}
x(t)&=&x(0)+v_x(0)t- \beta S^y(0)t
+\beta' S^x(0)(1-\cos\gamma B_0 t)
+ \beta' S^y(0) \sin\gamma B_0 t
\;,
\label{NMRa5}
\end{eqnarray}
where $\beta'={\beta}/{\gamma B_0 }$.

In the case of neutrons and for $B_0=1 \;\mathrm{T}$ and $B_1=300\; \mathrm{T}/\mathrm{m}$
we have $\beta\approx 2\times10^{-5}\;\mathrm{m}/\mathrm{s}$ and
$\beta'\approx 10^{-13} \;\mathrm{m}$
and it follows immediately from Equation~(\ref{NMRa5}) that the motion of the spin
has a negligible effect on the motion of the particles in the $x$-direction.

On the other hand, under the same conditions, it follows from Equations~(\ref{NMRa1c}) and~(\ref{NMRa2})
that $S^z(t)\approx S^z(0)$ and the equation of motion for the $z$-component of the velocity
becomes
\begin{eqnarray}
\frac{d v_z(t)}{d t}& =&\frac{\hbar\gamma B_1}{m}S^z(0)
\;,
\label{NMRa6}
\end{eqnarray}
yielding
\begin{eqnarray}
v_z(t)& =&v_z(0)+\frac{\hbar\gamma B_1}{m}S^z(0)t
\;.
\label{NMRa7}
\end{eqnarray}
In our simulations, $S^z(0)$ is a uniform random number in the range $[-1/2,+1/2]$.
Therefore, if $v_z(0)=0$, the values of $v_z(t^\ast)$ are also uniformly distributed
over the interval $[-v^\ast,+v^\ast]$.

Clearly, these elementary calculations yield results which are in excellent
agreement with the simulation data for $B_0=1 \;\mathrm{T}$.

\section[\appendixname~\thesection]{Numerical Solution of Equation~(\ref{NEWT3})}\label{VECTOR}
For any time step $\tau$, Equation~(\ref{NEWT3b}) can be solved in closed form.
In terms of the three components of the spin vector $\bS$, we have
\begin{equation}
\bS(t+\tau)=\left(\begin{array}{c}S^x(t+\tau)\\S_y(t+\tau)\\S^z(t+\tau)\\ \end{array}\right)
=\bR(\tau)
\left(\begin{array}{c}S^x(t)\\S_y(t)\\ S^z(t)\\ \end{array}\right)
=\bR(\tau)\bS(t)
\;,
\label{EQ8}
\end{equation}
where
\begin{equation}
\bR(\tau)=
\left(\begin{array}{ccc}
\uu^2+\vv^2{C}+\ww^2{C}& \uu\vv-\uu\vv{C}+\ww{S}& \uu\ww-\uu\ww{C}-\vv{S}\\
\uu\vv-\uu\vv{C}-\ww{S}&\vv^2+\uu^2{C}+\ww^2{C}&  \vv\ww-\vv\ww {C}+\uu{S}\\
\uu\ww-\uu\ww{C}+\vv{S}&\vv\ww-\vv\ww {C}-\uu{S}&\ww^2+\uu^2{C}+\vv^2{C}\\
\end{array}\right)
\;,
\label{EQ9}
\end{equation}
where
$\uu=\gamma B_x/\Omega$, $\vv=\gamma B_y/\Omega$ and $\ww=\gamma B_z/\Omega$,
${C}=\cos(\tau\Omega)$, ${S}=\sin(\tau\Omega)$ and $\Omega=|\gamma|(B_x^2+B_y^2+B_z^2)^{1/2}$.
The matrix $\bR(\tau)$ is orthogonal, implying that the integration
scheme does not change the length of $\bS$.

We integrate the equations of motion Equation~(\ref{NEWT3}) using
the velocity-Verlet algorithm~\cite{RAPA04}.
Initially, the positions and velocities are
normally distributed, centered  around $\bx=(0,0,0)$ and $\bv=(0,v_y,0)$
and with variances $\sigma^x$ and $\sigma^v$, respectively.

{\color{black}We only consider the case $0<y_0 < y_1$ and $y_0\le y\le y_1$.
According to Equation~(\ref{NEWT4}), at $t=0$, the force $\bF(\bx,t=0)=0$.
We choose a time step $\tau$ and repeat steps 1 to 5:
\begin{enumerate}
\item
$\bv \leftarrow \bv+\tau\bF/2m$,
\item
$\bx \leftarrow \bx+\tau\bv$,
\item
$\bS \leftarrow \bR(\tau)\bS$
\item
$\bF=\gamma\;B_1 S^z\be_z-\gamma\;B_1 S^x\be_x$,
\item
$\bv \leftarrow \bv+\tau\bF/2m$,
\end{enumerate}
for a number of time steps $N$.
}


As a curiosity, it may be of interest to mention that if the force $\bF$ is constant,
the Verlet scheme integrates the equation of motions {\sl exactly}.
On the other hand, Equations~(\ref{EQ8}) and~(\ref{EQ9}) integrate the torque
equation for spin {\sl exactly}.
Thus, it is only the combination of particle and spin motion that forces
us to integrate Equation~(\ref{NEWT3}) numerically.

\begin{figure}[!htp]

\includegraphics[width=0.47\hsize]{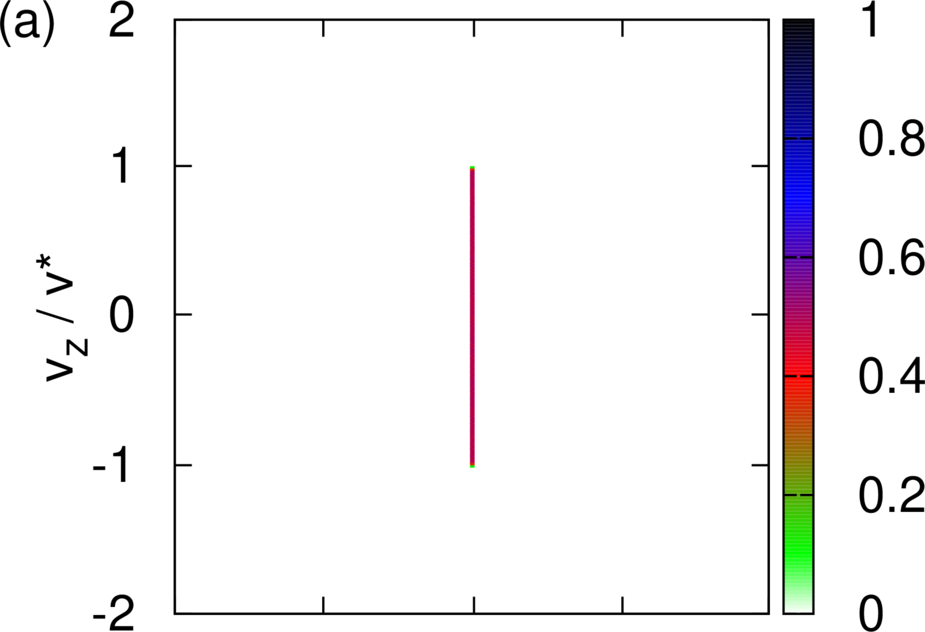}
\hspace{0.022\hsize}%
\includegraphics[width=0.47\hsize]{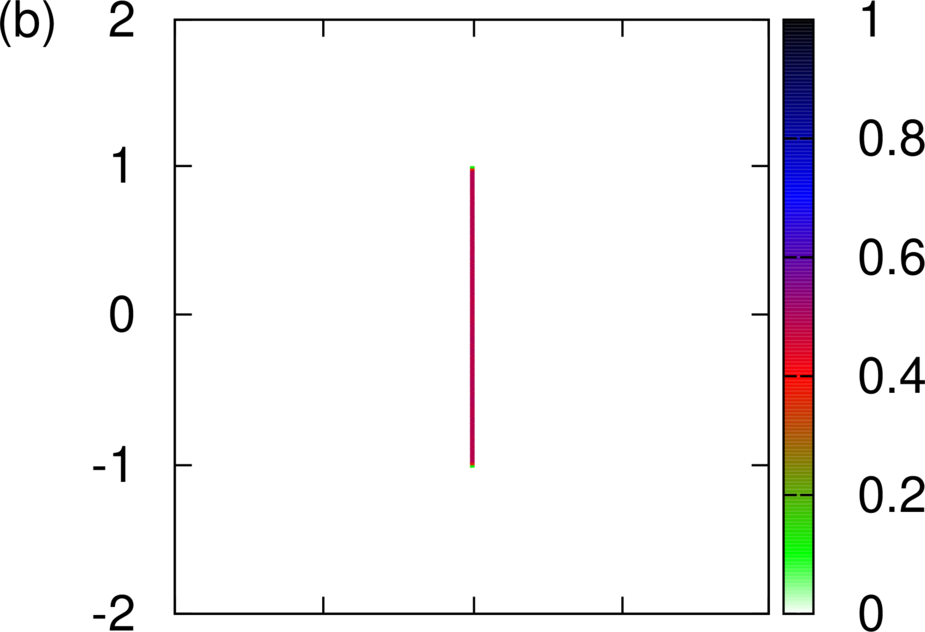}\\
\vspace{0.022\hsize}%
\includegraphics[width=0.47\hsize]{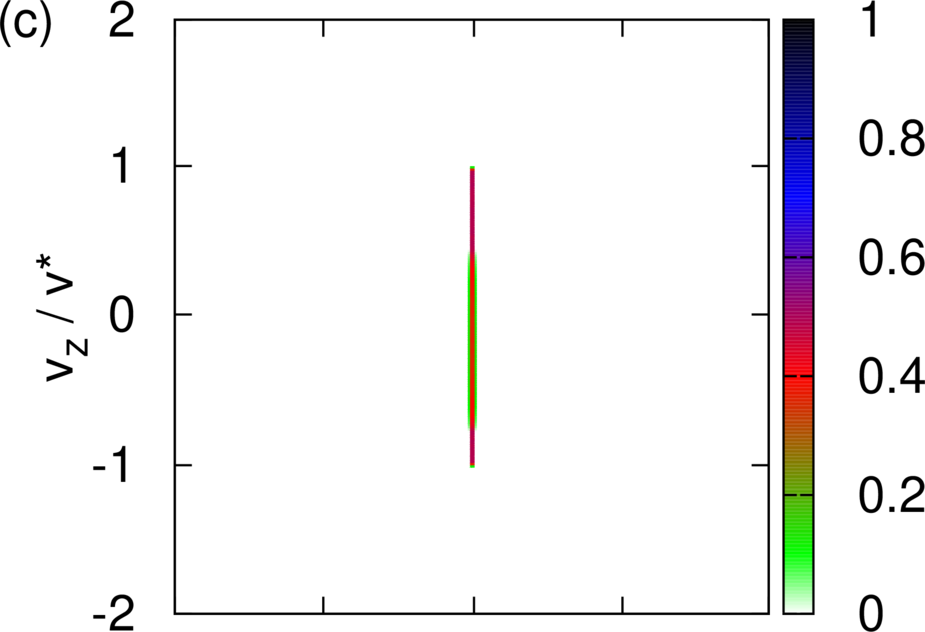}
\hspace{0.022\hsize}%
\includegraphics[width=0.47\hsize]{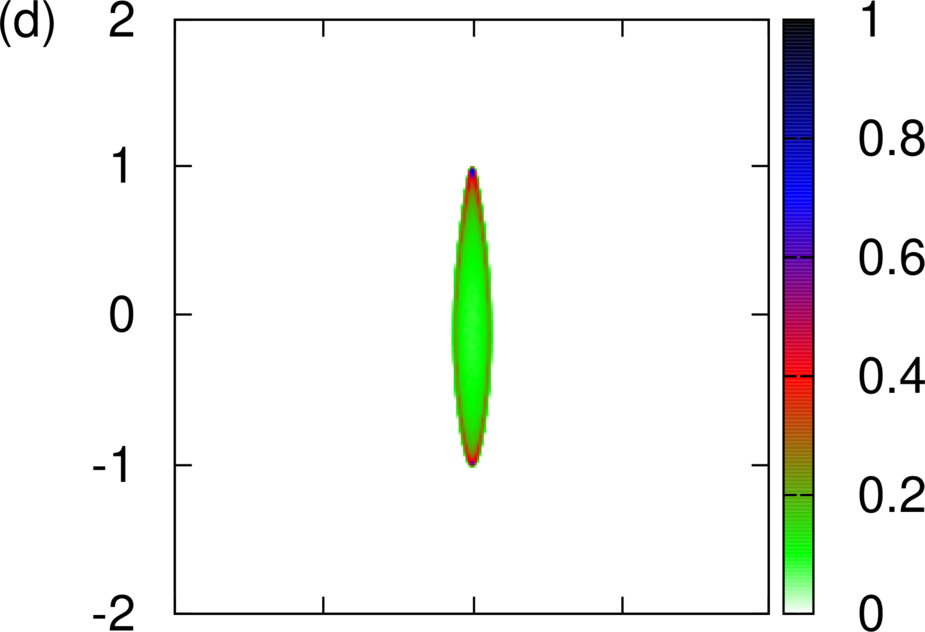}\\
\vspace{0.022\hsize}%
\includegraphics[width=0.47\hsize]{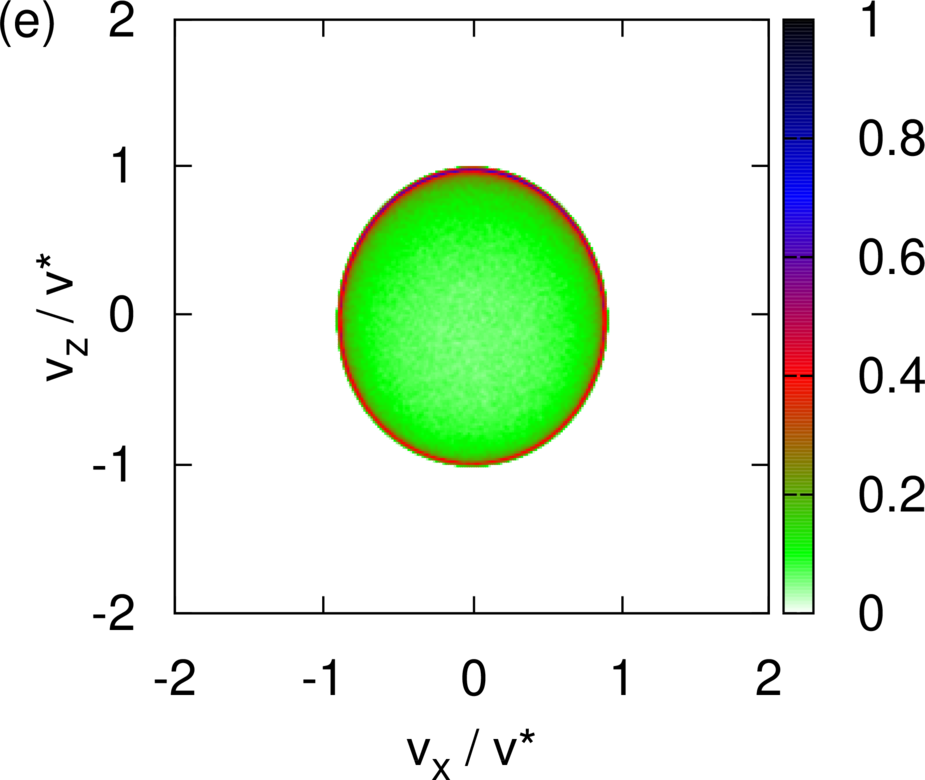}
\hspace{0.022\hsize}%
\includegraphics[width=0.47\hsize]{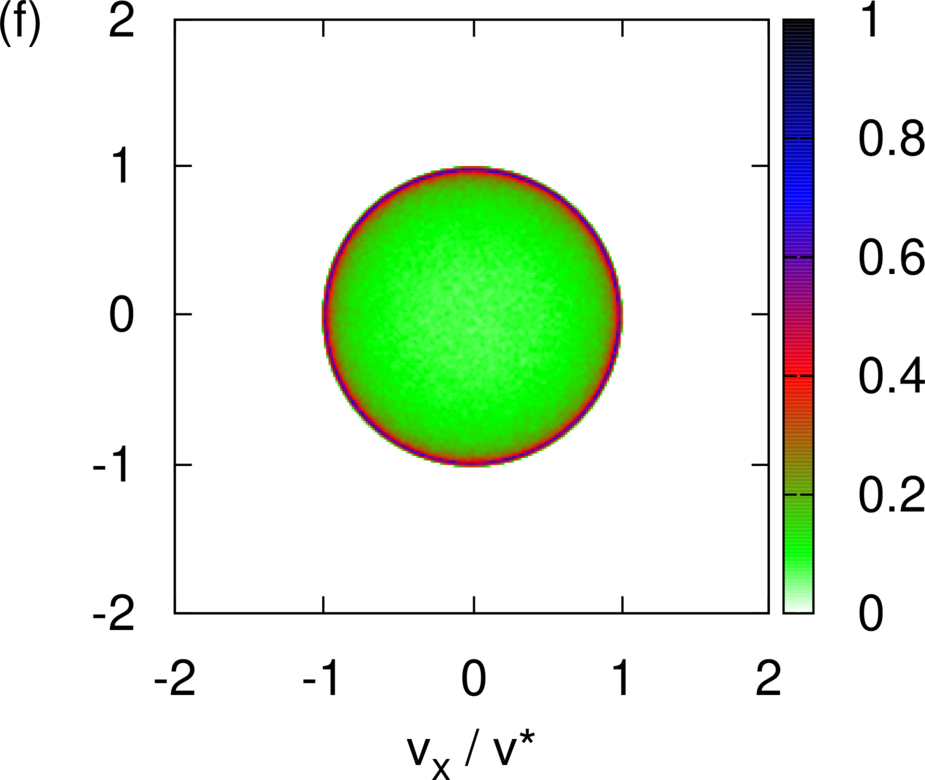}
\caption{{}
Histograms of the transverse velocity distribution obtained by
solving the classical equations of motion Equation~(\ref{NEWT3})
for the model parameters pertaining to {\color{black}imaginary silver particles}
and for different values of the uniform magnetic field $B_0$.
The initial magnetic moments distributed randomly (see text)
(\textbf{a}) $B_0=1\; \mathrm{T}$;
(\textbf{b}) $B_0=0.1\; \mathrm{T}$;
(\textbf{c}) $B_0=0.01\; \mathrm{T}$;
(\textbf{d}) $B_0=0.001\; \mathrm{T}$;
(\textbf{e})~$B_0=0.0001\; \mathrm{T}$;
(\textbf{f}) $B_0=0.00001\; \mathrm{T}$.
\label{NSIMAG0}
}
\end{figure}

\section[\appendixname~\thesection]{Newtonian Dynamics: {\color{black}Imaginary Silver Particles}}\label{NSIMAG}

{\color{black}Figure~\ref{NSIMAG0}c demonstrates that a uniform magnetic field of $B_0=0.01\; \mathrm{T}$}
is sufficiently large to yield a distribution of velocities
centered at $v_x=0$ and stretching from $v_z=-v^\ast$ to  $v_z=v^\ast$,
as expected for the classical, textbook model~\cite{STER22,FEYN65,BOHM51,BAYM74,BALL03}.

The transverse velocity distribution changes
drastically if $B_0$ decreases two and three orders of magnitude;
see Figure~\ref{NSIM0}d,e.

As $B_0$ vanishes, the transverse velocity distribution becomes a circular
disk of radius $v^\ast$, with the maximum located at the edges and the minimum at the center,
qualitatively similar to the case of neutrons.

\section[\appendixname~\thesection]{Quantum-Theoretical Model}\label{appQTM}
If we drop the term in $\sigma^x$ in Equation~(\ref{QTM2}),
the solution of the corresponding TDPE can be written as a product of unitary operators, each of which
can be worked out analytically.
{\color{black}
Writing ${\widehat H}= {\widetilde a}(p_x^2+p_z^2)-{\widetilde b}_0 \sigma^z+{\widetilde b}_1z \sigma^z$, we have
\begin{eqnarray}
U(t)&=&e^{-it{\widehat H}}
=e^{it{\widetilde b}_0\sigma^z}
e^{-it{\widetilde a}p_x^2}
e^{-it{\widetilde a}(p_z-{\widetilde b_1}t\sigma^z/2)^2} e^{it{\widetilde b_1}z\sigma^z}e^{-i{\widetilde a}{\widetilde b}^2t^3/12}
\;.
\label{PROP0}
\end{eqnarray}
}Computing the derivative of  $U(t)$ with respect to $t$, it readily follows that $U(t)$
satisfies $\partial_t U(t)={\widehat H}U(t)$.
In other words, if we drop the term in $\sigma^x$ in Equation~(\ref{QTM2}), the time-evolution operator
of the corresponding TDPE can be solved analytically; see also Appendix~\ref{TEX} for a more direct
proof of this fact.

\subsection[\appendixname~\thesubsection]{Momentum Representation}\label{MOM}
For the particular choice of the magnetic field given by Equation~(\ref{QTM1}),
it is advantageous for theoretical and numerical work to
write the TDPE Equation~(\ref{QTM2}) in the momentum representation~\cite{BALL03}.
We define Fourier transform pairs by
\begin{eqnarray}
\Psi(x,z,t)&=&\frac{1}{4\pi^2}\int_{-\infty}^{+\infty}\int_{-\infty}^{+\infty}  e^{i(k_xx+k_zz)}\,\Phi(k_x,k_z,t)
\;dk_x\,dk_z
\;,
\nonumber \\
\Phi(k_x,k_z,t)&=&\int_{-\infty}^{+\infty}\int_{-\infty}^{+\infty}  e^{-i(k_x x+k_z z)}\,\Psi(x,z,t) \;dx\,dz
\;,
\label{MOM0}
\end{eqnarray}
and introduce the operators
$q_x={i}\partial_{k_x}$ and
$q_z={i}\partial_{k_z}$.
Note that $[q_x,k_x]=[q_z,k_z]=i$ and that
the ``$i$'' in the definitions of $q_x$ and $q_z$
is not in the denominator as in the case
of the momentum operator in the coordinate representation.
The transformation to the momentum representation amounts to making the replacements
$x\leftrightarrow q_x$, $z\leftrightarrow q_z$,
$p_x\leftrightarrow k_x$, $p_z\leftrightarrow k_z$.
In the momentum representation, the TDPE Equation~(\ref{QTM2}) reads
\begin{eqnarray}
i\frac{\partial}{\partial t} |\Phi(t)\rangle=\left[
\frac{\hbar}{2m}\left(k_x^2+k_z^2\right)
-\frac{\gamma\, B_0}{2}\sigma^z
-\frac{i\gamma\, B_1}{2}\sigma^z \frac{\partial}{\partial k_z}
+\frac{i\gamma\, B_1}{2}\sigma^x \frac{\partial}{\partial k_x}
\right]
|\Phi(t)\rangle
\;,
\label{MOM2}
\end{eqnarray}
where
\begin{eqnarray}
\langle k_x,k_z|\Phi(t)\rangle=
\left(\begin{array}{ll} \Phi_{{+1}}(k_x,k_z,t) \\ \Phi_{{-1}}(k_x,k_z,t) \end{array}\right)
\;,
\label{MOM3}
\end{eqnarray}
is the two-component spinor in the momentum representation.

\subsection[\appendixname~\thesubsection]{Textbook Model}\label{TEX}
The TDPE Equation~(\ref{MOM2}) can be solved in closed form
if we ignore the term in Equation~(\ref{MOM2}) which is proportional to $\sigma^x$.
Then, the momentum $k_x$ and the projection of the spin in the $z$-direction are conserved; therefore, the motion in the $x$-direction is that of a free particle,
which we may omit from further considerations.
{\color{black}This simplified model is often used to discuss
the qualitative aspects of the SG experiment~\cite{FEYN65,BOHM51,BAYM74,Scully1987,Patil1998,BALL03},
but it does not comply with the Maxwell equations;}
however, as explained above, if the uniform magnetic field $B_0$ is sufficiently large,
the textbook model is an excellent approximation.

Not only is it instructive to {\color{black} derive the closed-form solution
of the TDPE of this simplified model, the solution itself is also of great help to validate simulation code.
Technically and for consistency, the treatment given below uses the momentum representation
(the treatment of which we have not found in the literature).}

The TDPE Equation~(\ref{MOM2}) of the textbook model separates into
two uncoupled first-order partial differential equations
\begin{eqnarray}
i\frac{\partial}{\partial t} \varphi_s(k_z,t)&=&\left[
\frac{\hbar k_z^2}{2m}
-\frac{s\gamma\, B_0}{2}
-\frac{is\gamma\, B_1}{2}\frac{\partial}{\partial k_z}
\right]
\varphi_s(k_z,t)
\;,
\label{TEX0}
\end{eqnarray}
where $s=\pm1$ denotes the eigenvalues of $\sigma^z$
and we omitted the constant term proportional to $k_x^2$.
We eliminate the term proportional to $B_0$ by a transformation
to a rotating frame.
Substituting $\varphi_s(k_z,t)=e^{ist\gamma\,B_0/2}\phi_s(k_z,t)$,
Equation~(\ref{TEX0}) becomes
\begin{eqnarray}
i\frac{\partial}{\partial t} \phi_s(k_z,t)&=&\left[
\frac{\hbar k_z^2}{2m}
-\frac{is\gamma\, B_1}{2}\frac{\partial}{\partial k_z}
\right]
\phi_s(k_z,t)
\;.
\label{TEX1}
\end{eqnarray}
Changing to new variables defined by $k_z=k'+gt'$ {\color{black} and} $t=t'$, we have
$\frac{\partial}{\partial k'}=\frac{\partial}{\partial k_z}$,
$\frac{\partial}{\partial t'}=\frac{\partial}{\partial t}+g\frac{\partial}{\partial k_z}$,
and Equation~(\ref{TEX1}) changes to
\begin{equation}
i\left[
\frac{\partial}{\partial t'}
+(\frac{s\gamma\, B_1}{2}-g)\frac{\partial}{\partial k'}\right]
\phi_s(k',t')=\frac{\hbar(k'+gt')^2}{2m}
\phi_s(k',t)
\;,
\label{TEX2}
\end{equation}
Choosing $g={s\gamma\, B_1}/{2}$, Equation~(\ref{TEX2}) simplifies to
\begin{eqnarray}
i\frac{\partial}{\partial t'} \phi_s(k',t')&=&\frac{\hbar(k'+gt')^2}{2m} \phi_s(k',t')
\;,
\label{TEX3}
\end{eqnarray}
the solution of which reads
\begin{eqnarray}
\phi_s(k',t')&=&e^{-i\hbar\,t'[k'^2+gt'+g^2t'^2/3]/2m} \phi_s(k',0)
\nonumber \\
&=&e^{-i{\hbar g^2t'^3}/{24m}} e^{-i{\hbar t'(k'+gt'/2)^2}/{2m}} \phi_s(k',0)
\;,
\label{TEX4}
\end{eqnarray}
or, in terms of the original coordinates,
\begin{equation}
\varphi_s(k_z,t)=e^{ist\gamma\,B_0/2}e^{-i{\hbar g^2t^3}/{24m}}
e^{-i{\hbar t(k_z-gt/2)^2}/{2m}} \varphi_s(k_z-gt,0)
\;.
\label{TEX5}
\end{equation}
It then follows that
\begin{eqnarray}
|\varphi_s(k_z+s\gamma\, B_1\,t/2,t)|^2&=&|\varphi_s(k_z,0)|^2\;,\;s=\pm1
\;,
\label{TEX6}
\end{eqnarray}
for any choice of the initial state $\varphi_s(k_z,0)$.
Equation~(\ref{TEX6}) tells us that as a function of time,
the probability density $|\varphi_s(k_z,t)|^2$ is the same as
the initial probability density, translated in momentum space  by $-s\gamma\, B_1\,t/2$.

Assuming that $\langle\varphi_s(k_z,0)|\varphi_s(k_z,0)\rangle=1$,
it follows from Equation~(\ref{TEX5}) that
\begin{eqnarray}
\langle k_z(t)\rangle_s&=&\langle\varphi_s(k_z,t)| k_z|\varphi_s(k_z,t)\rangle
\nonumber \\
&=&
\langle\varphi_s(k_z,t)| (k_z-gt)|\varphi_s(k,t)\rangle + gt\langle\varphi_s(k,t)|\varphi_s(k,t)\rangle
\nonumber \\
&=&\langle\varphi_s(k_z,0)| k_z|\varphi_s(k_z,0)\rangle + gt=\langle k_z(0)\rangle_s + gt
\nonumber \\
&=&
\langle k_z(0)\rangle_s + \frac{s\gamma\, B_1\,t}{2}
\;.
\label{TEX7}
\end{eqnarray}
Therefore, in the textbook case, the presence of a gradient in the magnetic field
causes the average momentum to linearly decrease ($s=+1$) or increase ($s=-1$)
if $\gamma<0$ (which is the case for neutrons or {\color{black}imaginary silver particles}).
Integrating Equation~(\ref{TEX7}) with respect to time yields
$\langle z(t)\rangle_s=\langle q_z(t)\rangle_s
=\langle z(0)\rangle_s+\langle k_z(0)\rangle_s\,t + s\gamma\, B_1\,t^2/4$
showing that the average position traces out a parabolic trajectory~\cite{BALL03}.

Writing Equation~(\ref{TEX7}) in terms of the quantum-theoretical velocity operator defined by
$v=\hbar k /m$, we have
$\langle v_z(t)\rangle_{\pm1}= \langle v_z(0)\rangle_{\pm1} \pm {\hbar\gamma\, B_1\,t}/{2m}$.
The classical mechanical expression $\pm \hbar\gamma B_1 S t/m$
for the change of the velocity due to the magnetic field gradient
matches the quantum-theoretical result if $S=1/2$, as mentioned earlier.


\subsection[\appendixname~\thesubsection]{Dimensionless Form}\label{DIME}

We define $\hbar k_x /m = v_0 v_x$,
$\hbar k_z /m = v_0 v_z$, and $t=t_0 \tau$
where $v_0$ and $t_0$ set the scale of the velocity and time, respectively.
In terms of these variables Equation~(\ref{MOM2}) reads
\begin{eqnarray}
i\frac{\partial}{\partial \tau} |\Phi(\tau)\rangle=\left[
\frac{mv_0^2t_0}{2\hbar}\left(v_x^2+v_z^2\right)
-\frac{\gamma\, B_0t_0}{2}\sigma^z
-\frac{i\hbar\gamma\, B_1t_0}{2mv_0}
\left(\sigma^z
\frac{\partial}{\partial v_z}
-
\sigma^x \frac{\partial}{\partial v_x}\right)
\right]
|\Phi(\tau)\rangle
\;.
\label{HAMI6}
\end{eqnarray}
We simplify the notation somewhat by introducing the (dimensionless) parameters
\begin{eqnarray}
a=\frac{mt_0v_0^2}{2\hbar}
\quad,\quad b= \frac{\hbar\gamma\, B_1t_0}{2mv_0}
\quad,\quad c= \frac{\gamma\, B_0t_0}{2}
\;,
\label{HAMI7}
\end{eqnarray}
and, at the risk of creating confusion, make the replacements
$\tau \rightarrow t$,
$v_x\rightarrow x$, $v_z\rightarrow z$, ${i}\frac{\partial}{\partial v_x} \rightarrow p_x$
and ${i}\frac{\partial}{\partial v_z} \rightarrow p_z$.
Then, Equation~(\ref{HAMI6}) becomes
\begin{equation}
i\frac{\partial}{\partial t} |\Phi(t)\rangle = 
\left[a\left(x^2+z^2\right)-c\sigma^z
-b\sigma^z p_z
+b\sigma^x p_x
\right]
|\Phi(t)\rangle
\;.
\label{HAMI8}
\end{equation}
We emphasize that from now on, whenever we
discuss the quantum-theoretical model,
$x$ and $z$ denote the dimensionless velocity
in the $x$- and $z$-direction, respectively.

In the case of neutrons, we use the parameters given in Equation~(\ref{VALU0})
and the corresponding values of $t_0=t^\star$ and $v_0=v^\star$ to find
\begin{eqnarray}
a&=&196540
\;,\;
b=-1
\;,\;
c=-185339 B_0\;\mathrm{T}^{-1}
\;,
\label{DIME3}
\end{eqnarray}
whereas for c (using Equation~(\ref{VALU1})), we find
\begin{eqnarray}
a&=&2.53618
\;,\;
b=-1
\;,\;
c=-8053.19 B_0\;\mathrm{T}^{-1}
\;.
\label{DIME4}
\end{eqnarray}
Comparing Eqs.~(\ref{DIME3}) and~(\ref{DIME4}), we may expect that
numerically solving the TDPE Equation~(\ref{HAMI8}) for the case of neutrons
is much more difficult than for the case of {\color{black}imaginary silver particles},
simply because in the former $a$ and $b$ differ by more than five orders of magnitude.

\subsection[\appendixname~\thesubsection]{Initial State}\label{INIT}

The initial two-component spinor in the momentum representation is given by
{\color{black}
\begin{eqnarray}
\langle x,z|\Phi(t=0)\rangle=
\left(\begin{array}{c}
\cos(\theta/2) e^{-i\alpha/2} \\ \sin(\theta/2) e^{+i\alpha/2} \end{array}\right)
G(x,z)
\;,
\label{INIT0}
\end{eqnarray}
}where $\theta$ controls the ratio and $\alpha$ controls the phases of
the spin-up and spin-down components, respectively.
The function $G(x,z)$ is taken to be
\begin{eqnarray}
G(x,z)&=&\frac{1}{2\pi\sigma^2}\exp\left(-\frac{x^2+z^2}{2\sigma^2}\right)
\;,
\label{INIT1}
\end{eqnarray}
the standard Gaussian distribution with variance $\sigma$ and centered around $(0,0)$.

In practice, we either use uniform random numbers
{\color{black} to determine $\cos(\theta/2)$} and $\alpha$
or we set $\alpha=\pi/4$ and choose $\theta$ from the set
$\{\pi/4\,,\,-\pi/4\,,\,0\,,\, \pi/2\}$, corresponding
to the spin states
$(|\uparrow\rangle + |\downarrow\rangle)/\sqrt{2}$,
$(|\uparrow\rangle - |\downarrow\rangle)/\sqrt{2}$,
$|\uparrow\rangle$, and
$|\downarrow\rangle$, respectively.

\subsection[\appendixname~\thesubsection]{Simulation Method}\label{TECH}

In practice, we solve Equation~(\ref{HAMI8}) by means of the Chebyshev polynomial algorithm~\cite{TALE84,RAED03y}.
Disregarding the discretization of the $p_x$ and $p_z$,
this algorithm has the virtue that it yields numerical results with close to machine precision,
for any time $t$~\cite{TALE84,RAED03y}.
Technical details about this TDPE solver can be found in Appendix~\ref{TECHa}.

For the reasons explained in Appendix~\ref{TECHa},
simulating the neutron experiment~\cite{HAME75} is prohibitively costly.
A simple way out of this conundrum is to ``make the SG magnet shorter'' by a factor of ten.
Then $t_0=t^\star/10$, $v_0=v^\star/10$ and, according to Equation~(\ref{HAMI7}),
$a$ is reduced by a factor of thousand and $b$ is left unchanged.

A simulation with $t_0=t^\ast/10$ and $v_0=v^\ast/10$,
using a grid of $2^{15}\times2^{15}$ points requires about 200 GiB of memory
and finishes in about 8 hours, using a compute node with
two AMD EPYC 7742, $2\times64$ cores, 2.25 GHz processors.
Although it would be technically straightforward to use more nodes by
extending the code to use the MPI communication protocol,
we believe that this extension would not bring much additional insight.
The main point here is that we can perform simulations in which the
neutrons travel a macroscopic distance and the dimensions of the apparatus are also macroscopic.

It may be of interest to mention that in practice, a product-formula approach~\cite{RAED87,Hsu2011} using
the decomposition in terms of the exact propagators in Equation~(\ref{PROP0}) for the $x$ and $z$ components
fails, simply because of the large disparity between the coefficients $a$ and $b$ (in the case of neutrons).

\subsection[\appendixname~\thesubsection]{TDPE Solver: Technical Aspects}\label{TECHa}

We discretize the $x$ and $z$ variables by using a square regular grid.
The mesh size $\delta$ should be small enough to
support (i) an accurate representation of terms with the first derivatives and (ii)
accurately resolve the dependence of the ``potential'' $a(x^2+z^2)$
on $x$ and $z$.
As $|b|=1$, condition (i) is automatically satisfied if (ii) is satisfied.
For the neutron case, the fact that $a$ is more than five orders of magnitude
larger than $|b|$ makes it difficult to satisfy condition (ii).

Ideally, for the numerical solution of Equation~(\ref{MOM2}),
we would like to take $t_0=t^\ast$ and $v_0=v^\ast$ as the scales of time and velocity,
respectively. We do so in the case of {\color{black}imaginary silver particles}.
Unfortunately, for the case of neutrons
an accurate numerical solution of Equation~(\ref{MOM2}) for $t_0=t^\ast$ and $v_0=v^\ast$
requires a currently prohibitive amount of memory and CPU time.

To appreciate the difficulty of satisfying condition (ii) it is sufficient to focus on the $x$-dependence
and consider the case $B_0=0$.
We assume $\delta\ll1$ in the following.
Initially, the wave packet is concentrated around $-m\delta \le x\le m\delta$
where $m$ is a small (compared to the grid size) integer.
The potential of two neighboring grid points in the vicinity of $x=0$ differs by
the amount $a(x\pm\delta)^2-ax^2\approx \pm a\delta^2(2m\pm1)$.
On the other hand, with our choice of dimensionless units, the relevant range
of the dimensionless velocities $x$ and $z$ is approximately $[-1,1]$.
If, in the course of time, (part of) the wave packet is concentrated
around say $x=1$, the potential of two neighboring grid points
differs by the amount $a(x\pm\delta)^2-ax^2\approx \pm 2a\delta$.
In order to represent the smoothly changing potential $ax^2$
on a grid, we must require $2a\delta\ll1$.
Otherwise, when the wave moves towards $x=\pm1$, it
will encounter a potential that changes in big steps
and its dynamics will no longer resemble that of a wave
propagating in a continuum.
As explained below, the number of grid points in one direction that we use is
of the order $2^{15}$, yielding a grid size of $\delta=4/2^{15}=2^{-13}\approx0.00012$.
For neutrons $a=196540$, $2a\delta\approx48$, which is much too large.
In contrast, for {\color{black}imaginary silver particles}  $a=2.53618$, $2a\delta\approx0.0006$, which is definitely small enough; therefore, in the case of neutrons, we are forced to reduce the size of the simulation problem.

\subsection[\appendixname~\thesubsection]{Quantum Dynamics: {\color{black}Imaginary Silver Particles}}\label{QSIMAG}

Unlike in the case of neutrons in which memory requirements limited the
TDPE integration time to $t^\star/10$, the parameters for {\color{black}imaginary silver particles}
are such that there are no such limitations. Thus, in this case $v_0=v^\ast$.
The value of $B_0$ at which the two spots changes into a circular shape
is hard to predict without actually performing the simulation.

\begin{figure}[!htp]

\includegraphics[width=0.47\hsize]{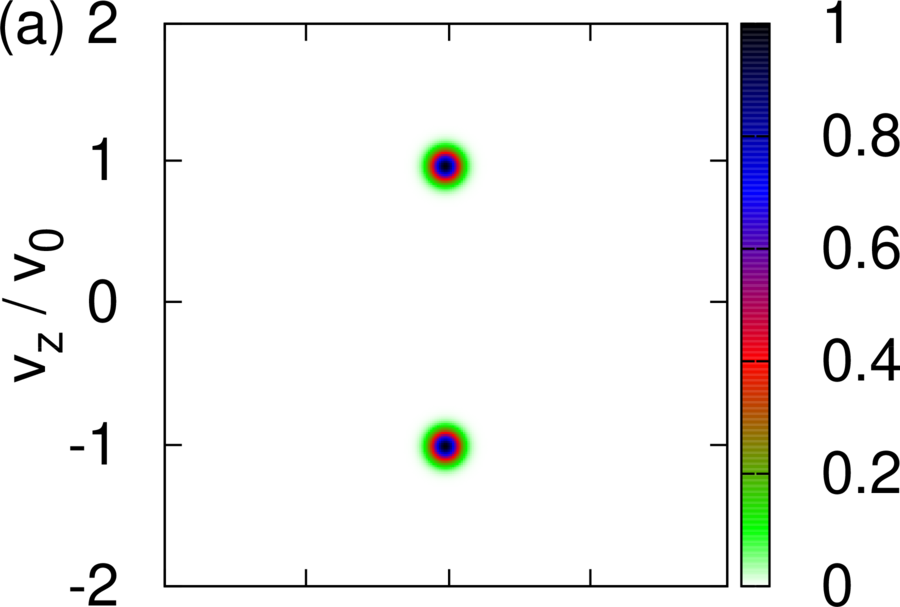}
\hspace{0.022\hsize}%
\includegraphics[width=0.47\hsize]{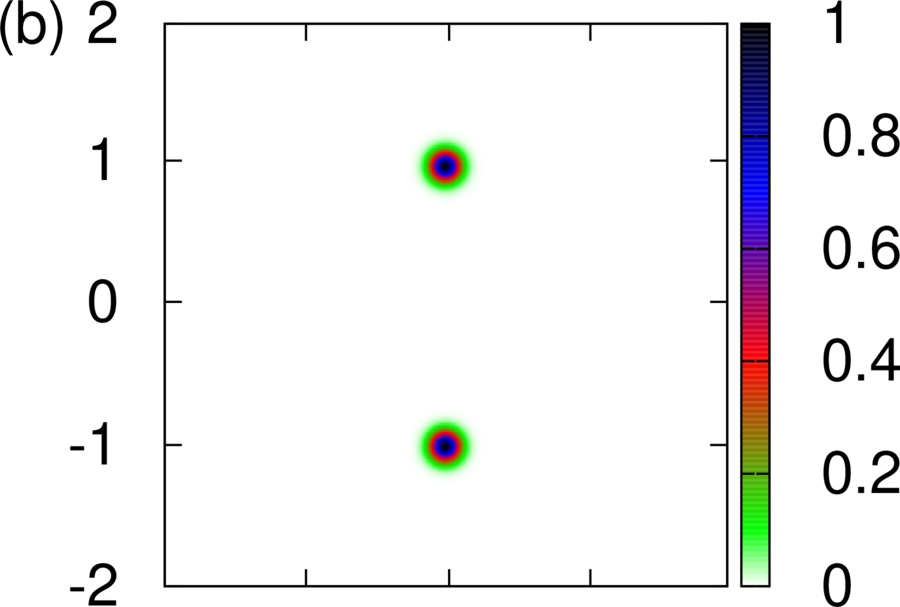}\\
\vspace{0.022\hsize}%
\includegraphics[width=0.47\hsize]{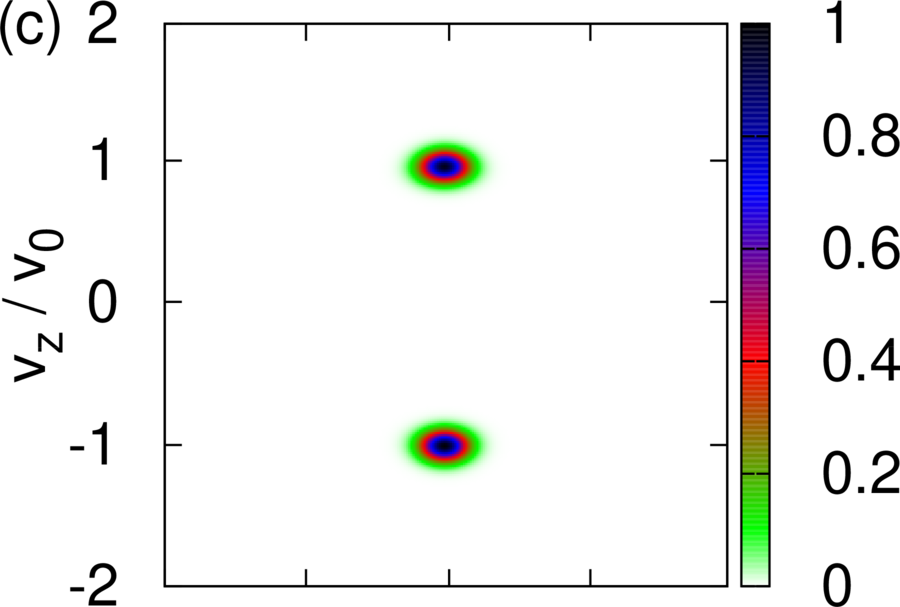}
\hspace{0.022\hsize}%
\includegraphics[width=0.47\hsize]{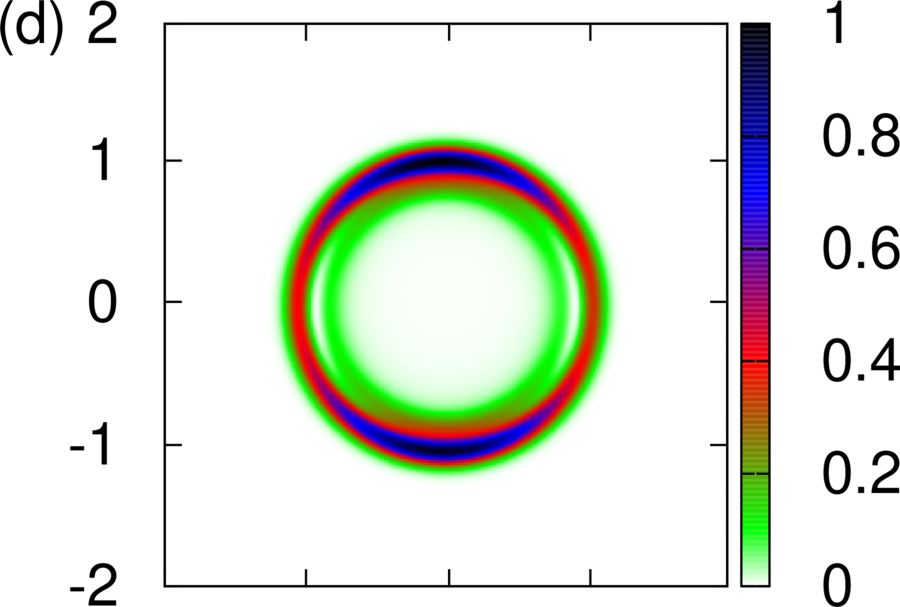}\\
\vspace{0.022\hsize}%
\includegraphics[width=0.47\hsize]{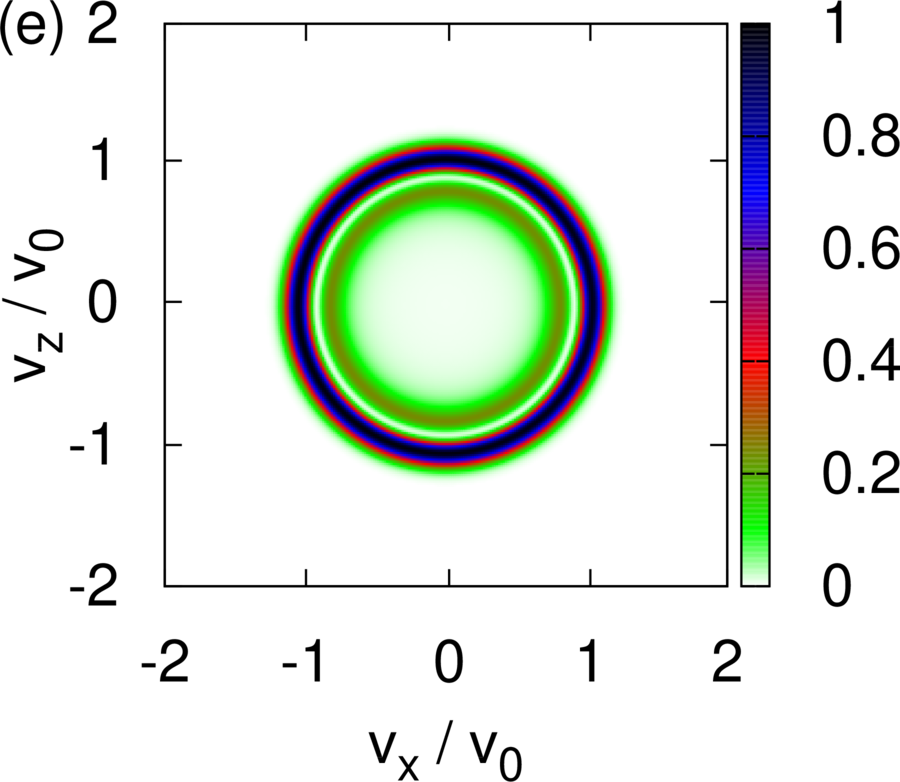}
\hspace{0.022\hsize}%
\includegraphics[width=0.47\hsize]{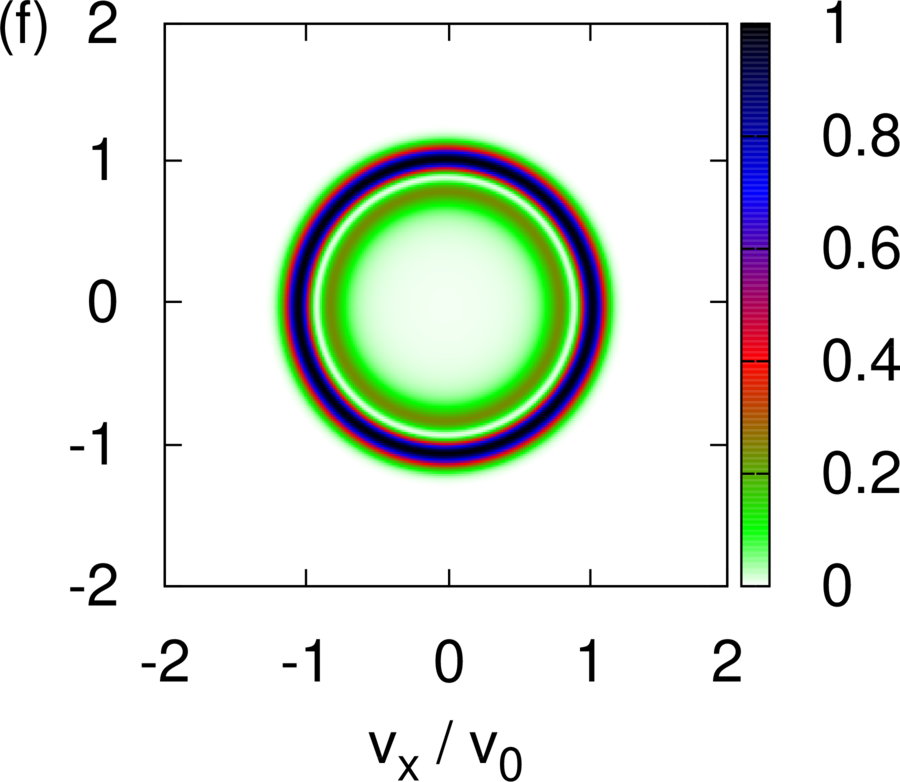}
\caption{{}
Probability distributions {\color{black}$|\langle v_x,v_z|\Phi(t^\ast)\rangle|^2$}
of the transverse velocity distribution obtained by
solving the TDPE Equation~(\ref{HAMI8}) with the initial state given
by Equation~(\ref{INIT0})
and for the model parameters pertaining to {\color{black}imaginary silver particles} ($v_0=v^\ast$).
Initially, the variance (dimensionless) $\sigma=0.1$ and the spin state is $(|\uparrow\rangle + |\downarrow\rangle)/\sqrt{2}$.
(\textbf{a}) $B_0=1\; \mathrm{T}$;
(\textbf{b}) $B_0=0.1\; \mathrm{T}$;
(\textbf{c})~$B_0=0.01\; \mathrm{T}$;
(\textbf{d}) $B_0=0.001\; \mathrm{T}$;
(\textbf{e}) $B_0=0.0001\; \mathrm{T}$;
(\textbf{f}) $B_0=0.00001\; \mathrm{T}$.
\label{QSIMAG0}
}
\end{figure}






\clearpage
\end{document}